\documentclass{aa}
\usepackage{graphicx}

\usepackage{xcolor}
\usepackage{ulem}
\usepackage{comment}
\usepackage{natbib}
\usepackage{amssymb}
\usepackage{scrextend} 
\usepackage{placeins} 
\usepackage{hyperref}
\hypersetup{
    colorlinks=true,
    linkcolor=blue,
    filecolor=magenta,      
    urlcolor=blue,
    citecolor=blue,
    }
\usepackage{multirow}
\usepackage{txfonts}
\begin{document} 

   \title{Galactic runaway O and Be stars found using \textit{Gaia} DR3 \thanks{The complete versions of Tables~\ref{Tab:GOSC_Runaways} and \ref{Tab:BeSS_Runaways} are available at the CDS via anonymous ftp to \href{https://cdsarc.cds.unistra.fr}{\tt cdsarc.cds.unistra.fr} (\href{ftp://130.79.128.5}{\tt 130.79.128.5}) or via \url{https://cdsarc.cds.unistra.fr/viz-bin/cat/J/A+A/vol/page}}}

   \author{M. Carretero-Castrillo
          \inst{1}
          \and M. Rib\'o \inst{1}\fnmsep\thanks{Serra H\'unter Fellow}
          \and J. M. Paredes \inst{1}
          }

   \institute{Departament de Física Quàntica i Astrofísica, Institut de Ciències del Cosmos (ICCUB), Universitat de Barcelona (IEEC-UB),
              Martí i Franquès 1, 08028 Barcelona, Spain\\
              \email{mcarretero@fqa.ub.edu}
             }

   \date{Received April 6, 2023; accepted July 10, 2023}

 
  \abstract
   {A relevant fraction of massive stars are runaway stars. These stars move with a significant peculiar velocity with respect to their environment.}
   {We aim to discover and characterize the population of massive and early-type runaway stars in the GOSC and BeSS catalogs using \textit{Gaia} DR3 astrometric data.}
   {We present a two-dimensional method in the velocity space to discover runaway stars as those that deviate significantly from the velocity distribution of field stars. Field stars are considered to follow the Galactic rotation curve.}
   {We found 106 O runaway stars, 42 of which were not previously identified as runaways.
   We found 69 Be runaway stars, 47 of which were not previously identified as runaways.
   The dispersion of runaway stars is a few times higher in $Z$ and $b$ than that of field stars. This is explained by the ejections they underwent when they became runaways.
   The percentage of runaways is 25.4\% for O-type stars, and it is 5.2\% for Be-type stars.
   In addition, we conducted simulations in three dimensions for our catalogs. They revealed that these percentages could increase to $\sim$30\% and $\sim$6.7\%, respectively.
   Our runaway stars include seven X-ray binaries and one gamma-ray binary.
   Moreover, we obtain velocity dispersions of $\sim$5~km~s$^{-1}$ perpendicular to the Galactic plane for O- and Be-type field stars. These values increase in the Galactic plane to $\sim$7~km~s$^{-1}$ for O-type stars due to uncertainties and to $\sim$9~km~s$^{-1}$ for Be-type stars due to Galactic velocity diffusion.}
   {The excellent \textit{Gaia} DR3 astrometric data have allowed us to identify a significant number of O-type and Be-type runaways in the GOSC and BeSS catalogs.
   The higher percentages and higher velocities found for O-type compared to Be-type runaways underline that the dynamical ejection scenario is more likely than the binary supernova scenario. Our results open the door to identifying new high-energy systems among our runaways by conducting detailed studies.}


   \maketitle
%

\section{Introduction}

Massive early-type OB stars are the most luminous stars in the Milky Way. They are relevant for the study of the kinematics and dynamics of the young stellar populations, the metallicity enrichment in the Galaxy, and the feedback in the interstellar medium through gamma-ray bursts and supernova explosions, among many other topics (see, e.g., \citealt{Vanbeveren1998,Woosley2006}). O stars are characterized by powerful stellar winds and short life-spans (<10$^7$~yr), while B stars have longer life-spans (up to $\sim$10$^8$~yr). Be stars are a particular type of B stars that display Balmer emission lines and an excess of infrared emission because they present circumstellar envelopes in the form of decretion disks \citep{Slettebak1988,Rivinius2013}. A relevant feature of OB stars is that they are mostly found in binaries, while at least 70\% of them form interacting binaries (\citealt{Chini2012}; \citealt{Sana2012}; \citealt{MoeStefano2017}). After stellar evolution, these systems can form high-mass X-ray binaries \citep{Lewin2006}, gamma-ray binaries \citep{Dubus2013}, millisecond pulsar systems, and double neutron stars \citep{Heuvel2007}, which allow the study of nonthermal processes.

On the other hand, more than 30\% of the O stars and about 5--10\% of the B stars are considered to be runaway stars. These stars have a high peculiar velocity with respect to their environment (\citealt{Blaauw1961}; \citealt{Stone1979}; \citealt{Boubert2018}). There are two possible scenarios to explain the existence of runaway stars: (1) the dynamical ejection scenario (DES), where a close three- or four-body interaction in the core of a dense cluster ejects a massive star \citep{Poveda1967}, and (2) the binary supernova scenario (BSS), where the mass loss in the supernova (SN) explosion of the most evolved star in a binary system induces either a runaway velocity to the remaining binary system, which now contains a compact object, or disrupts the binary system and induces runaway velocities to both the remaining massive OB star and the remaining compact object \citep{Blaauw1961}. Later works explored the idea that massive runaway stars could host compact companions (\citealt{BekensteinBowers1974}; \citealt{Stone1982}; \citealt{Oijen1989}). There is now observational evidence of runaway X-ray binaries (see, e.g., \citealt{Heuvel2000}). In addition, three out of the ten known gamma-ray binaries are runaways: LS~5039, PSR~B1259$-$63, and 1FGL~J1018.6$-$5856 (\citealt{Marcote2018} and references therein). Therefore, runaway massive stars can be part of close binary systems. In a sample of O runaway stars, \cite{Chini2012} obtained a binary fraction of 69\%.

Several works that searched for massive runaway stars can be found in the literature, from early efforts based on radial and space velocities from the literature (\citealt{CruzGonz1974}; \citealt{Stone1979}) and those using \textit{Hipparcos} proper motions (\citealt{Hoogerwerf2001}; \citealt{Mdzinarishvili2004}; \citealt{Wit2005}; \citealt{Tetzlaff2011}) to recent publications using \textit{Gaia} DR1, DR2, and EDR3 data (\citealt{MA2018}; \citealt{Boubert2018}; \citealt{Kobulnicky2022}). Not only is there diversity in the input data, but also in the runaway detection method. Some authors based the criterion for detecting runaways on the three-dimensional (3D) peculiar velocity of the stars above a velocity threshold, which is usually in a range from 25 to 40~km~s$^{-1}$ (\citealt{Blaauw1961}, \citealt{CruzGonz1974}, \citealt{Hoogerwerf2001}). Other authors complemented this criterion with that of the distance of the stars above or below the Galactic plane, setting  $|Z| > 0.25$--0.50~kpc as the limit (\citealt{Wit2005}, \citealt{Tetzlaff2011}). Because precise radial velocities and distances are lacking, other authors used a two-dimensional (2D) proper-motion-only method, which allows detecting about half of the existing runaways \citep{MA2018}. \cite{Kobulnicky2022} used a method in which a star is considered as runaway when it has a 2D peculiar velocity above 25~km~s$^{-1}$ with respect to its environment, assuming a Galactic rotation curve. This diversity of input data and methods translates into a diversity of results.

Our aim in this work is to search for new massive O and Be runaway stars, some of which could be (new) high-mass X-ray or gamma-ray binaries. In particular, the discovery of new gamma-ray binaries may shed light on some of the unresolved questions in these systems \citep{Dubus2013,Dubus2017}. To this end, we started to search for runaway massive stars using O-star catalogs and \textit{Gaia} DR2 data using a 2D velocity method  \citep{Ayan2019}, as well as O- and Be-star catalogs and \textit{Gaia} EDR3 data \citep{Carretero-Castrillo2023}. In this work, we continue this project, but use \textit{Gaia} DR3 data and improve several aspects of the method.

This paper is organized as follows. In Sect.~\ref{Sec:Catalogs} we describe the different catalogs we used. In Sect.~\ref{Sec:Cross-match} we explain the cross match of the massive star catalogs with \textit{Gaia} DR3, the quality cuts we applied to the data, and the properties of the resulting catalogs. In Sect.~\ref{Sec:Methodology} we describe our method, how we computed the velocities and related uncertainties, and the criteria we used to classify runaway stars. In Sect.~\ref{sec:results} we present the results we obtained for the field and runaway stars. In Sect.~\ref{Sec:Discussion} we discuss the velocity dispersion of the field stars, the spatial distribution of the runaways, compare our results with previous works, discuss the percentage of runaways as a function of the spectral type, and briefly comment on the X-ray and gamma-ray binaries we found. We conclude by summarizing our main findings in Sect.~\ref{Sec:Conclusions}.

\section{Catalogs} \label{Sec:Catalogs}

\subsection{Gaia}

\textit{Gaia} is an astrometric mission of the European Space Agency (ESA) that was launched in 2013. The main objective of this spacecraft is to measure the 3D spatial and velocity distribution of stars in the Galaxy with unprecedented precision. It also provides astrophysical properties of stars, including surface gravity and effective temperature. Thus, the mission allows mapping and revealing the formation, structure, and evolution of the Milky Way in the magnitude range $G=3$--21 \citep{GaiaMission}.

Over time, \textit{Gaia} data are made public through data releases that are opened to the scientific community. The latest release, \textit{Gaia} DR3 \citep{GaiaDR3}, was made public on 13 June 2022 and is used in this work. This release, which contains more than $1.8\times10^9$ objects, constitutes a breakthrough in terms of quality, quantity and variety of astrophysical data. In particular, it represents the largest data set of all sky spectrophotometry and radial velocities. However, although radial velocities for $33.8\times10^6$ stars are included, these velocities are scarce and sometimes inaccurate for the massive stars of our interest because of the priors that were used \citep[see, e.g.,][]{Drew2022}. The astrometric data of DR3, namely positions $(\alpha,\delta)$, proper motions $\left(\mu_{\alpha^*}=\mu_\alpha\cos{\delta}, \mu_{\delta}\right),$ and parallaxes ($\varpi$), are the same as those already included in \textit{Gaia} EDR3 \citep{GaiaEDR3}, with reference epoch 2016.0.

We point out, however, that the EDR3 parallax uncertainties were underestimated, as explained in \cite{Fabricius2021} and also noted by other authors (e.g., \citealt{MA2021}). In addition, \cite{MA2022} provides an alternative method for deriving parallax zeropoints to the methods proposed by \cite{Lindegren2021}, which is better suited for bright stars. Therefore, to obtain accurate and precise parallaxes we followed the procedure presented in \cite{MA2021} and \cite{MA2022}. We note that this procedure requires $G>6$ and five- or six-parameter solutions in \textit{Gaia}. However, we also used these restrictions when we applied our own quality cuts (see Sect.~\ref{Sec:Cross-match}). To apply this procedure, first we computed corrected parallaxes using the equation $\varpi_{\rm c}=\varpi-Z_{\rm EDR3}$, where $\varpi$ are the original \textit{Gaia} parallaxes and $Z_{\rm EDR3}$ is the parallax zeropoint, which depends on magnitude, color, and ecliptic latitude, using Eq.~(4) of \cite{MA2022}. Second, we computed the so-called external parallax uncertainties using their formula $\sigma_{\rm ext}=\sqrt{k^2\sigma_{\rm int}^2+\sigma_{\rm s}^2}$, where $k$ is a magnitude-dependent constant, $\sigma_{\rm int}$ are the original \textit{Gaia} parallax uncertainties, and $\sigma_{\rm s}$ is the systematic uncertainty for individual parallaxes.

\subsection{Galactic O-Star Catalog}

The Galactic O-Star Catalog (GOSC) is an ongoing project that collects the most accurate information of Galactic O-type stars \citep{GOSC}. In this catalog, special attention is paid to the correct spectral classification of the stars. To achieve this goal, this team started the Galactic O-Star Spectroscopic Survey (GOSSS; \citealt{GOSSS}). GOSSS contains blue-violet spectra of Galactic O stars with intermediate spectral resolution ($R \sim$ 2500) and high signal-to-noise ratio ($\geq$~300). Accordingly, from version three onwards, GOSC contains GOSSS spectral types. GOSC provides J2000 coordinates, spectral types, luminosity classes, and related clusters and associations, among many other items. The astrometric precision is 0.001~s in right ascension and 0.01\arcsec\ in declination. The $V$ magnitude ranges from $\sim$2 to $\sim$14.

In this work, we use the main catalog and supplement 2 of GOSC, whose current versions (v4.2) contain 611 O stars and 32 BA stars, respectively, with a total of 643 stars. This catalog contains precise spectral classifications for all stars, allowing us to select only OB-type stars. Since for this catalog we are only interested in early-type stars up to spectral type B1, we removed two A0 stars from the catalog. We also eliminated 37 sources that were identified with the multiple star system flag because their astrometric data might not be reliable enough (not only in GOSC, but presumably also in \textit{Gaia}). Consequently, the catalog contains 604 stars.

\subsection{Be Star Spectra}

The Be Star Spectra (BeSS) Database \citep{BeSS} provides a catalog of Be stars that is as complete as possible. It assembles spectra obtained by professional and amateur astronomers. The current version contains 2330 Be stars, Herbig Ae/Be stars, and B[e] supergiants of the Galaxy and the Large and Small Magellanic Clouds (LMC and SMC). The catalog provides J2000 coordinates of each star and the apparent magnitude $V$ and spectral type for most of them. The precision is 0.01~s in right ascension and 0.01\arcsec\ in declination. For this catalog, the visual magnitude $V$ ranges from $\sim$2 to $\sim$20. The information is collected from the Simbad database and the literature, when this is available. Therefore, although this catalog may contain observational biases, it includes current information of a large number of Be stars, which makes it useful for our purposes. We eliminated all stars with spectral type A or that were flagged as Herbig stars from BeSS, which represented 88 and 55 stars, respectively. We also removed 126 and 215 stars with coordinates closer than 10\degr\ to the centers of the LMC and SMC, respectively \citep{vanderMarel2001}, because we are only interested in Galactic runaways. As a result, the catalog contained 1881 stars (some stars fulfilled more than one requirement for their exclusion). We note that a few Oe stars are present in both the GOSC and BeSS catalogs, but we did not remove them from any of them despite the overlap because we wished to analyze both catalogs independently.

\section{Cross match of the catalogs} \label{Sec:Cross-match}

\subsection{Cross match of GOSC and BeSS with \textit{Gaia} DR3}

We first cross-matched the GOSC catalog with \textit{Gaia} DR3 using a radius of 1\arcsec\ within the \textit{Gaia} cross match tool of the ESA \textit{Gaia} archive\footnote{\href{https://gea.esac.esa.int/archive}{gea.esac.esa.int/archive}\label{refnote}}. We computed the differences between the GOSC and \textit{Gaia} DR3 coordinates and found standard deviations of $\sim$0\farcs07 in both coordinates. We restricted the cross-match radius to 0\farcs5 to avoid false positives, which represents $\text{about seven}$ standard deviations. Of the 604 GOSC stars, we obtained 600 cross matches.

In the case of BeSS, we proceeded in a similar way and found standard deviations of $\sim$0\farcs17. However, the distributions looked Gaussian in the center, but showed long tails. We therefore decided to use a cross-match radius of 1\farcs5 for BeSS, which represents $\text{about nine}$ standard deviations and includes correct matches of stars in the long tails.  We obtained 1863 cross matches for the 1881 BeSS
stars. Because the BeSS coordinates might not be very accurate, we compared the BeSS $V$ and \textit{Gaia} DR3 $G$ magnitudes to search for possible misidentifications. We inspected all cases where $|V-G|>1$ and verified that the \textit{Gaia} source identifier or \texttt{source\_id} provided by the Simbad database was the same as the one provided by the \textit{Gaia} cross-match tool. This was always the case, except for a single star, Cl*~NGC~6871~BP~2, which had $|V-G|> 4$, for which we manually reassigned the correct \textit{Gaia} \texttt{source\_id}.

The cross matches yielded stars with more than one \textit{Gaia} DR3 \texttt{source\_id}. For the GOSC catalog, we obtained two stars with two different \textit{Gaia} DR3 \texttt{source\_id}, while for the BeSS catalog, we obtained 55 stars. Of these, we only retained the \texttt{source\_ids} whose coordinates were closer to those of the respective catalogs, which in all cases had similar magnitudes. After the cross match, we had 598 GOSC stars (6 stars were lost) and 1808 BeSS stars (73 stars were lost).

We then applied different quality cuts to the resulting catalogs to ensure a good quality of the astrometric data. These quality cuts are summarized below and are presented in Table~\ref{Tab:Cuts}, together with the number of stars affected by each of them. The first cut corresponds to the opposite case of the previous paragraph: the same \textit{Gaia} DR3 \texttt{source\_id} attributed to two different stars.  We found 6 stars in pairs with the same \texttt{source\_id} for the GOSC catalog and 2 stars for the BeSS catalog. We removed all of them because the accuracy of the astrometric solution would be compromised. We then applied the quality cuts recommended by \cite{Lindegren2021} for \textit{Gaia} data, as well as some additional cuts used by other authors (e.g., \citealt{BailerJones2015}). In this way, we rejected stars without a five- and six-parameter solution, a $G$ magnitude lower than 6, fewer than ten visibility periods, an any apparent fractional parallax error (ratio of the corrected external parallax uncertainty to the corrected  parallax, $\sigma_{\rm ext}/\varpi_{\rm c}$) higher than 0.2, a negative parallax, and a renormalized unit weight error (RUWE) parameter $>1.35$, which indicates a poor astrometric solution. We used a slightly lower value of RUWE than 1.4 that is mentioned in the \textit{Gaia} data release documentation\footref{refnote} to ensure good-quality data.
As we explain in Sect.~\ref{Sec:Methodology}, we also excluded stars with galactocentric distances smaller than 5~kpc, which led to an additional loss of 6 GOSC stars and 1 BeSS star. After applying all these cuts, we had what we call the GOSC-\textit{Gaia}~DR3 catalog, which contains 417 stars, and the BeSS-\textit{Gaia}~DR3 catalog, which contains 1335 stars (we note that 12 of these are Oe stars that are also present in the  GOSC-\textit{Gaia}~DR3 catalog).

\subsection{Properties of GOSC-\textit{Gaia}~DR3 and BeSS-\textit{Gaia}~DR3}

\begin{figure}[t!]
    \centering
    \includegraphics[width=\hsize]{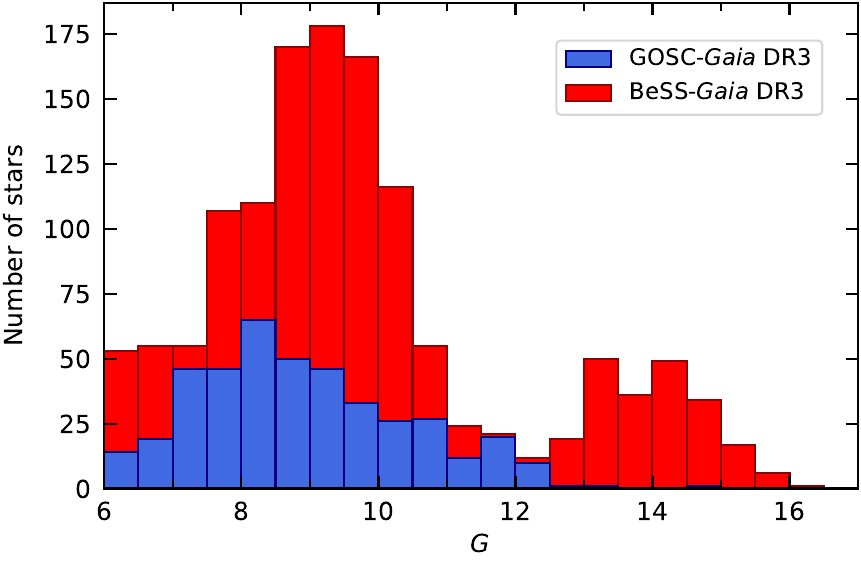}
      \caption{Histogram of the $G$ magnitudes for GOSC-\textit{Gaia}~DR3 stars in blue and BeSS-\textit{Gaia}~DR3 stars in red, with a bin size of 0.5 magnitudes.
              }
    \label{Fig:Mag}
   \end{figure}

Figure~\ref{Fig:Mag} shows the distribution of $G$ magnitudes for the stars of the GOSC-\textit{Gaia}~DR3 catalog in blue and the BeSS-\textit{Gaia}~DR3 catalog in red. The distributions are represented using a bin size of 0.5 magnitudes. The lower limit of the distribution corresponds to the $G>6$ quality cut, and the upper limit reaches $\sim$15 and $\sim$17 for the O- and Be-type stars, respectively. The magnitudes of the GOSC-\textit{Gaia}~DR3 stars have a maximum around 8--9, and the BeSS-\textit{Gaia}~DR3 stars have maxima of $\sim$9 and $\sim$14. The upper end of the first maximum is due to the observational limit of the instruments used for the majority of the Be star studies. The second maximum corresponds to a detection of Be stars with larger telescopes or very low-resolution surveys \citep{BeSS}.

To compute distances $d$ to the stars (from the Sun), we inverted the corrected \textit{Gaia} parallaxes $\varpi_{\rm c}$ discussed in Sect.~\ref{Sec:Catalogs}. This procedure introduces significant biases and provides asymmetric distance uncertainties, particularly for fractional parallax uncertainties above our quality-cut limit of 0.2 \citep{BailerJones2015}. In addition, we are aware that the quality cut in fractional parallax uncertainty biases the sample toward nearby and bright objects \citep{Luri2018}. However, our original sample is already formed by bright and relatively nearby objects, and we only lost an additional $\sim$4\% of the stars by using this quality cut in addition to the others discussed above. In contrast, this allowed us to obtain a cleaner sample. We show the distribution of the distances from the Sun for both catalogs in Fig. 2. The stars are mainly located up to 3~kpc from the Sun, but the distributions extend up to 9~kpc. The nearest GOSC-\textit{Gaia}~DR3 stars are at $\sim$0.7~kpc because of the $G>6$ magnitude limit, which prevents the presence of nearby bright stars. The BeSS-\textit{Gaia}~DR3 stars are typically closer. Their distribution starts at $\sim$0.1~kpc because they are fainter and, considering the whole sample, less affected by the magnitude limit.

    \begin{figure}[t!]
    \centering
    \includegraphics[width=\hsize]{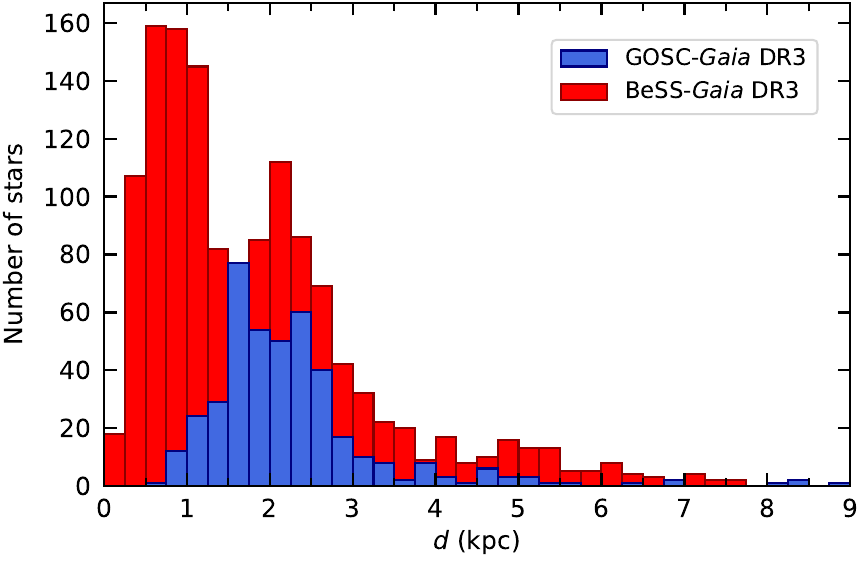}
    \caption{Histogram of the distances to the Sun of GOSC-\textit{Gaia}~DR3 stars in blue and of BeSS-\textit{Gaia}~DR3 stars in red, with a bin size of 0.25~kpc.}
    \label{Fig:Dist}
   \end{figure}

\begin{figure}[t!]
    \centering
    \includegraphics[width=\hsize]{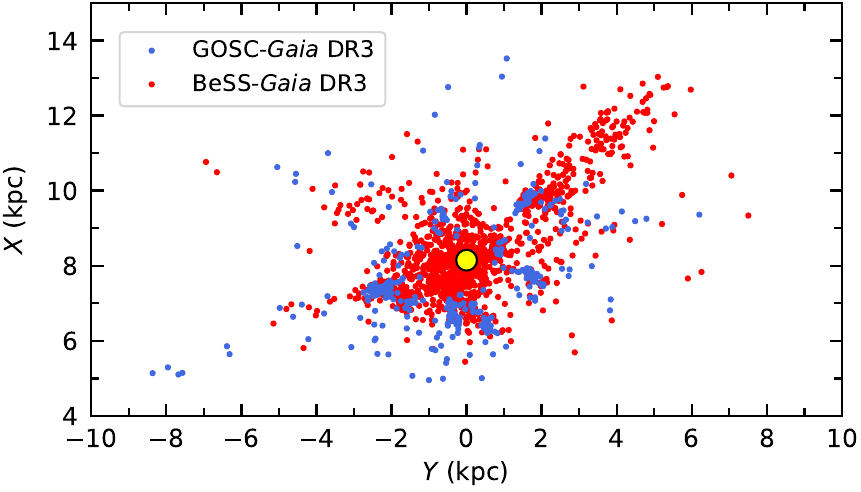}
      \caption{Galactocentric $XY$ coordinates. The coordinates of the GOSC-\textit{Gaia}~DR3 stars are shown in blue, and those of the BeSS-\textit{Gaia}~DR3 stars are plotted in red. The position of the Sun is marked with a yellow circle at $\left(X, Y\right) = \left(8.15,0\right)$~kpc. The Galactic center is located at $\left(0,0\right)$~kpc.}
              \label{Fig:XYGal}
   \end{figure}

To display the positions of the stars in our catalogs in galactocentric Cartesian coordinates, we used the convention in which the $X$-axis starts at the Galactic center and points toward the Sun,  the $Y$-axis is perpendicular to it and defined positive toward the direction of the Galactic rotation for the Sun, and the $Z$-axis is perpendicular to the $X$- and $Y$-axes. To compute these coordinates, we first computed the $(l,b)$ coordinates using the two equatorial coordinates of the North Galactic Pole (NGP), $\left(\alpha_{\text{NGP}}, \delta_{\text{NGP}}\right)$, and the position angle of the North Celestial Pole with respect to the great semicircle passing through the zero Galactic longitude and the NGP, $\theta_0$, all taken from \cite{Reid2004}. Second, we used the obtained $(l,b)$ angular coordinates and considered a galactocentric solar radius of $R_{\sun} = 8.15$~kpc \citep{Reid2019} to compute the galactocentric Cartesian coordinates $XYZ$ and the galactocentric distance $R_*$. We show in Fig.~\ref{Fig:XYGal} the $XY$ galactocentric coordinates for the stars in our catalogs, where the Galactic center is located at $\left(0,0\right)$~kpc. The absence of nearby GOSC-\textit{Gaia}~DR3 stars due to the cut in magnitude is clearly visible compared to the distribution of BeSS-\textit{Gaia}~DR3 stars. O-type stars, which are younger than Be stars, are expected to follow the spiral arms of the Milky Way better. Comparing Fig.~\ref{Fig:XYGal} with Fig.~2 of \cite{Xu2021}, we can see that our O-type stars approximately trace the Perseus, the Local, and the Sagittarius-Carina arms. In addition, a comparison of Fig.~\ref{Fig:XYGal} with Fig.~5 of \cite{Pantaleoni2021} reveals that O-type stars also lies within the two interarm structures they discovered: the \object{Vela~OB1} association and the Cepheus spur. The overdensity of BeSS-\textit{Gaia}~DR3 stars in the direction from the Sun toward $XY = (13,5)$~kpc is due to low-resolution surveys of Be stars in that particular direction, and it corresponds to the second magnitude peak in Fig.~\ref{Fig:Mag}. Although not shown here, most of the stars in our catalogs are located close to the Galactic plane, with mean and standard deviations of $Z=-0.00\pm0.11$~kpc for the GOSC-\textit{Gaia}~DR3 stars and $Z=-0.01\pm0.19$~kpc for the BeSS-\textit{Gaia}~DR3 stars.

\section{Search for runaways: Method} \label{Sec:Methodology}

Our goal is to search for runaway stars among the stars in our catalogs. Runaway stars move with a significant peculiar velocity with respect to the mean Galactic rotation, which is the one followed by field stars. To reach this goal, we have to compute and analyze the velocities of the stars, handle the propagation of uncertainties, and define a criterion to classify the stars as runaways.

\subsection{Velocities with respect to the regional standard of rest} \label{Sec:RSRvels}

\begin{figure}[t]
    \centering
    \includegraphics[width=0.9\hsize]{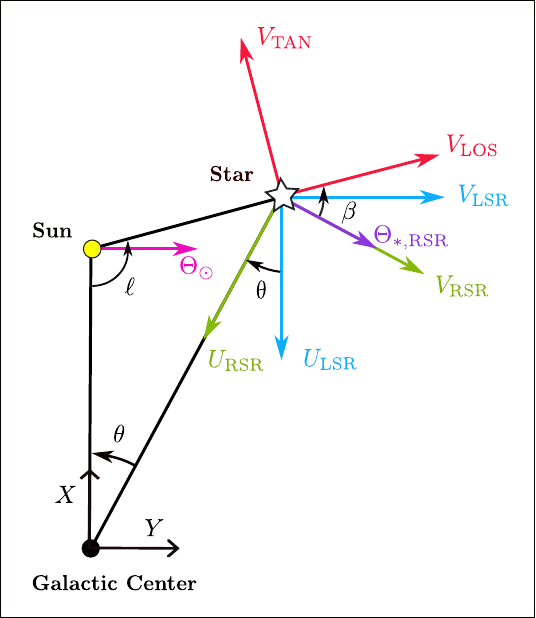}
    \caption{Reference systems used in this work, depicted in the $XY$ plane. The Sun and the Galactic center are indicated with a yellow and black circle, respectively. The white star represents a given star in our catalogs. $l$ is the Galactic longitude, and $\theta$ and $\beta$ are defined in the text. $\Theta_\sun$ is the circular rotation velocity at the position of the Sun, and $\Theta_{*,{\rm RSR}}$ is the rotational velocity at the position of the star. LSR and RSR velocities are shown in light blue and green, respectively. The tangential velocity $V_\text{TAN}$ and the line-of-sight velocity $V_\text{LOS}$ are shown in red. The third velocity components of the LSR and the RSR systems, $W_\text{LSR}$ and $W_\text{RSR}$, are not shown in the figure because they are perpendicular to this plane.}
      \label{Fig:ReferenceSystems}
\end{figure}

The Galactic velocity components for any given star are defined as follows: $U$ is positive toward the Galactic center, $V$ is positive in the direction of the Galactic rotation, and $W$ is positive toward the NGP. To compute the velocities of the stars, we used the different reference systems presented in Fig.~\ref{Fig:ReferenceSystems}. First, we computed the Galactic velocity components with respect to the local standard of rest (LSR), $\left(U_{\text{LSR}}, V_{\text{LSR}}, \text{and } W_{\text{LSR}}\right)$. To do this, we applied the transformations of \cite{Johnson&Soderblom} and added the solar motion with respect to the LSR $\left(U_{\sun}, V_{\sun}, W_{\sun}\right)$. In these equations, we took the equatorial coordinates $\left(\alpha, \delta\right)$ and the proper motions $\left(\mu_{\alpha^*}=\mu_\alpha\cos{\delta}, \mu_{\delta}\right)$ from \textit{Gaia} DR3, our corrected \textit{Gaia} DR3 parallaxes $\varpi_{\rm c}$ (see Sect.~\ref{Sec:Cross-match}), and $\left(\alpha_{\text{NGP}}, \delta_{\text{NGP}}\right)$ and $\theta_0$ from \cite{Reid2004}. We used the solar motion values provided by the A5 fit of \cite{Reid2019}: $U_{\sun}=10.8 \pm 1.8$, $V_{\sun}= 13.6 \pm 6.8$, and $W_{\sun}=7.6 \pm 1.0$~km~s$^{-1}$.

The transformations of \cite{Johnson&Soderblom} require the radial velocity $\rho$ , which is not available in \textit{Gaia}~DR3 data for most of our stars (only $\sim$10\% of the GOSC-\textit{Gaia}~DR3 stars, and $\sim$3\% of the BeSS-\textit{Gaia}~DR3 stars have \textit{Gaia} radial velocities). To circumvent this problem, we assigned the theoretical radial velocity to each start that it should have if it were moving in its regional standard of rest (RSR) according to a Galactic rotation curve with a velocity $\Theta_{*,{\rm RSR}}$ at galactocentric distance $R_*$. We used the Galactic rotation curve provided by the A5 fit of \cite{Reid2019}: circular rotation velocity at the position of the Sun $\Theta_\sun = 236 \pm 7$~km~s$^{-1}$ , and a distance from the Sun to the Galactic center of $R_\sun = 8.15 \pm 0.15$~kpc. Because these authors did not provide the rotation curve slope, we applied a linear fit to their data listed in Table~4, Column~3, which are shown in the bottom panel of their Fig.~11. However, the Galactic rotation curve deviates from the linear behavior for galactocentric distances smaller than $5$~kpc, therefore we discarded a few stars with $R_*<5$~kpc, as already mentioned in Sect.~\ref{Sec:Cross-match}. Our linear fit for galactocentric distances greater than $5$~kpc, fixing $\Theta=\Theta_\sun$ for $R=R_\sun$, provides a slope of $d\Theta/dR = -0.4 \pm 0.4$~km~s$^{-1}$~kpc$^{-1}$. With $R_\sun$, $\Theta_\sun$, $d\Theta/dR,$ and $R_*$ we computed $\Theta_{*,{\rm RSR}}$ for each star and derived the theoretical radial velocity considering the Galactic coordinates $(l,b)$, the angle $\theta$ of the star shown in Fig.~\ref{Fig:ReferenceSystems}, $\Theta_\sun$ and $U_{\sun}$, $V_{\sun}$, and $W_{\sun}$. This allowed us to determine $U_{\text{LSR}}$, $V_{\text{LSR}}$, and $W_{\text{LSR}}$.

The next step was to determine the velocities of the star with respect to the RSR. The velocities were computed as follows:
\begin{equation}
    \begin{cases}
      U_\text{RSR} = U_\text{LSR}\cos{\theta}-V_\text{LSR}\sin{\theta} - \Theta_{\sun}\sin{\theta}\\
      V_\text{RSR} = U_\text{LSR}\sin{\theta}+V_\text{LSR}\cos{\theta} + \Theta_{\sun}\cos{\theta}-\Theta_{{*}\text{,RSR}}\\
      W_\text{RSR} = W_\text{LSR},
    \end{cases} 
\end{equation}

The distributions of the GOSC-\textit{Gaia}~DR3 and BeSS-\textit{Gaia}~DR3 velocities $\left(U_{\text{RSR}}, V_{\text{RSR}}, W_{\text{RSR}}\right)$ are presented and discussed in Appendix~\ref{Sec:App_velocities}.

At this point, we defined a new reference system by applying a rotation to the Galactic plane velocities $U_\text{RSR}$ and $V_\text{RSR}$ (see Fig.~\ref{Fig:ReferenceSystems}) in such a way that the unknown radial velocity mainly affects the line-of-sight component, $V_\text{LOS}$, and does not affect the other component, $V_\text{TAN}$. This allowed us to minimize the influence of the theoretical radial velocity. The components of the new system were computed as follows:
\begin{equation}
    \label{Eq:NewSystem}
    \begin{cases}
      V_\text{LOS} = - V_\text{RSR}\sin{\beta} - U_\text{RSR}\cos{\beta} \\
      V_\text{TAN} = V_\text{RSR}\cos{\beta} - U_\text{RSR}\sin{\beta}\\ 
      W_\text{RSR} = W_\text{LSR},
    \end{cases} 
\end{equation}
where $V_{\text{TAN}}$ is the tangential velocity contained in the plane of the sky and parallel to the Galactic plane, $V_{\text{LOS}}$ is the line-of-sight velocity, and $W_{\text{RSR}}$ did not change. In these equations, $\beta = l + \theta - 90$. From then on, we left $V_{\text{LOS}}$ and worked with the 2D system composed of $V_{\text{TAN}}$ and $W_{\text{RSR}}$.

\begin{figure}[t!]
    \centering
    \includegraphics[width=\hsize]{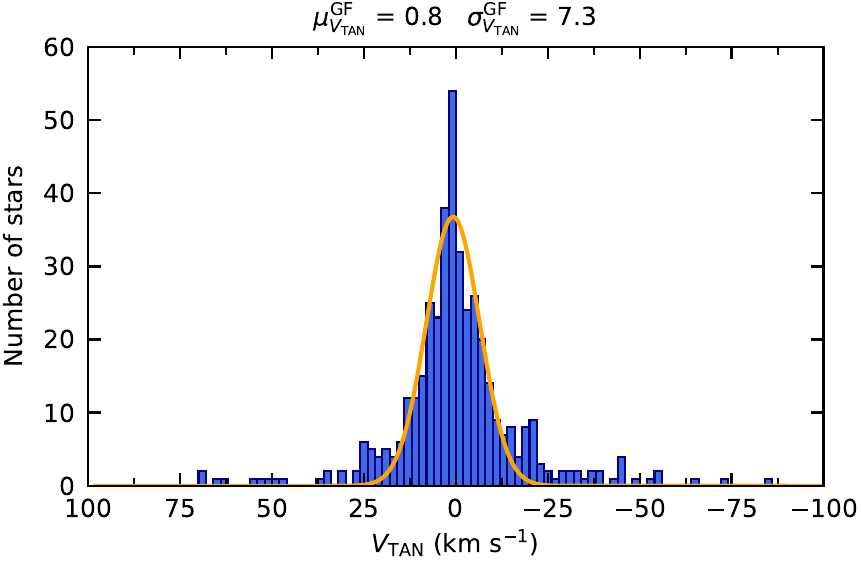}\vspace{3 mm}
    \includegraphics[width=\hsize]{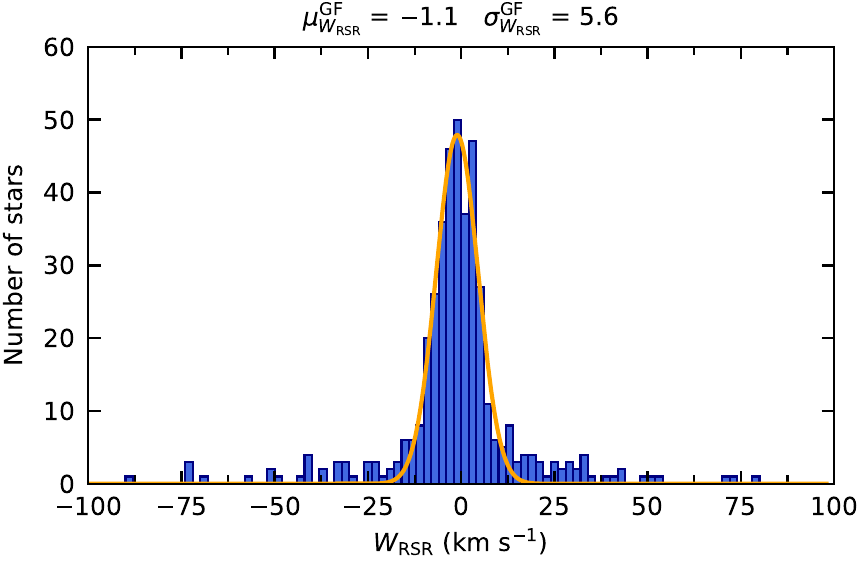}
    \caption{Distributions of the $V_{\text{TAN}}$ and $W_{\text{RSR}}$ velocities for the GOSC-\textit{Gaia}~DR3 stars. The histograms have a bin size of 2~km~s$^{-1}$. The orange lines represent Gaussian functions (GF) fitted to the data, whose means and standard deviations in km~s$^{-1}$ are quoted above the panels. The abscissa ranges have been limited to $\pm 100$~km~s$^{-1}$.
    \textit{Top}: histogram of the $V_{\text{TAN}}$ velocities.
    \textit{Bottom}: histogram of the $W_{\text{RSR}}$ velocities.}
      \label{Fig:GOSC_VTAN_WRSR}
\end{figure}

\begin{figure}[t]
    \centering
    \includegraphics[width=\hsize]{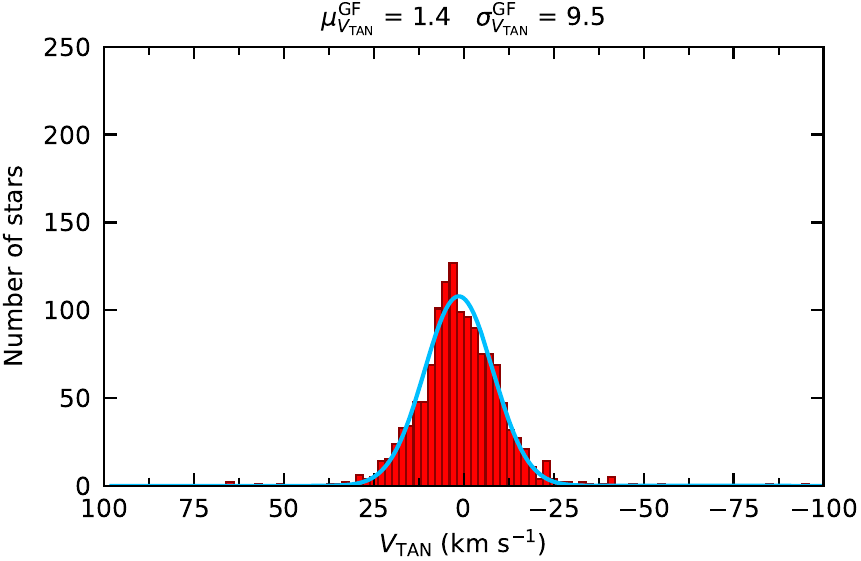}\vspace{3 mm}
    \includegraphics[width=\hsize]{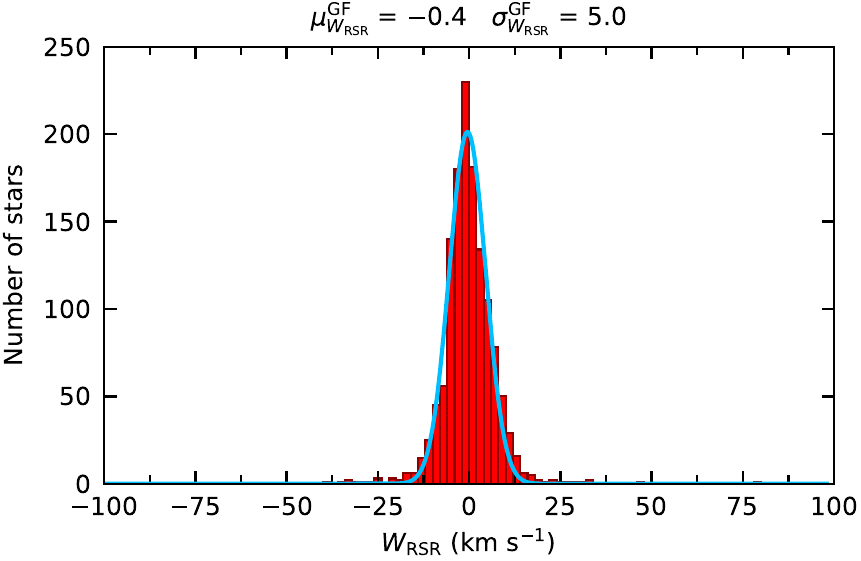}
    \caption{Same as Fig.~\ref{Fig:GOSC_VTAN_WRSR} but for the BeSS-\textit{Gaia}~DR3 stars and with the Gaussian functions fitted to the data shown in blue.}
      \label{Fig:BeSS_VTAN_WRSR}
\end{figure}

The histograms of the GOSC-\textit{Gaia}~DR3 velocities $V_{\text{TAN}}$ and $W_{\text{RSR}}$ with a bin size of 2~km~s$^{-1}$ are shown in Fig.~\ref{Fig:GOSC_VTAN_WRSR} (the $V_{\text{TAN}}$ axis is inverted according to the geometry of our reference system). The figures are limited to the $\pm 100$~km~s$^{-1}$ abscissa range for better visualization. Two and three stars are beyond this range for the $V_\text{TAN}$ and the $W_\text{RSR}$ components, respectively, and their highest absolute velocities are $\lvert V_{\text{TAN}}\rvert = 179.7$ and $\lvert W_{\text{RSR}}\rvert= 129.1$~km~s$^{-1}$. The histograms show velocities clustered around zero that approximately follow Gaussian functions (superimposed in the figures), and some stars clearly deviate from this behavior. This is expected for field stars and for runaway stars, respectively. We note that stars with $V_{\text{TAN}}=0$ follow the Galactic rotation curve, while stars with $W_{\text{RSR}}=0$ only move in the Galactic plane. The Gaussian fits shown in the figure were computed considering all the stars and are thus affected by the runaway stars, which are identified in the following sections. The histograms of the BeSS-\textit{Gaia}~DR3 velocities $V_{\text{TAN}}$ and $W_{\text{RSR}}$ with a bin size of 2~km~s$^{-1}$ are shown in Fig.~\ref{Fig:BeSS_VTAN_WRSR}. The figures are limited to the $\pm 100$~km~s$^{-1}$ abscissa range for better visualization. Only one star lies outside this range in the $V_{\text{TAN}}$ component, whose absolute velocity is $\lvert V_{\text{TAN}}\rvert = 131.1$~km~s$^{-1}$.

\subsection{Propagation of uncertainties}\label{Sec:Uncertainties}

In all our computations, the sources of uncertainty are the five astrometric parameters $\left(\alpha, \delta, \mu_{\alpha^*}, \mu_{\delta}, \text{and }\varpi_{\text{c}}\right)$, the solar velocities $\left(U_{\sun}, V_{\sun}, \text{and }W_{\sun}\right)$, the distance from the Sun to the Galactic center $R_\sun$, the circular rotation velocity at the position of the Sun $\Theta_\sun$, and the slope of the Galactic rotation curve $d\Theta/dR$. We call $X$ any of the parameters of the sources of uncertainty $\left(\alpha, \delta, \mu_{\alpha^*}, \mu_{\delta}, \varpi_{\text{c}}, U_{\sun}, V_{\sun}, W_{\sun}, R_\sun, \Theta_\sun,\text{and } d\Theta/dR\right)$, and $Y$  the functions in which the uncertainties are propagated: positions, angles, and velocities. The error on a function $Y$ was computed as $\sqrt{J_XCJ^{T}_X}$, where $J_X$ is the Jacobian matrix of all the parameters $X$, and $C$ is the covariance matrix, which only contains diagonal elements in the
case of independent variables\footnote{We note that the five astrometric parameters from \textit{Gaia} are correlated. However, the use of $\varpi_{\text{c}}$ prevents us from using the original correlations from \textit{Gaia}. Treating the other four parameters as correlated or independent made no difference to the computed uncertainties in our case.}. These diagonal elements are the uncertainties of each of our parameters $\sigma_X$. This method assumes symmetric uncertainties, as is the case here.

\begin{figure}[t!]
    \centering
    \includegraphics[width=\hsize]{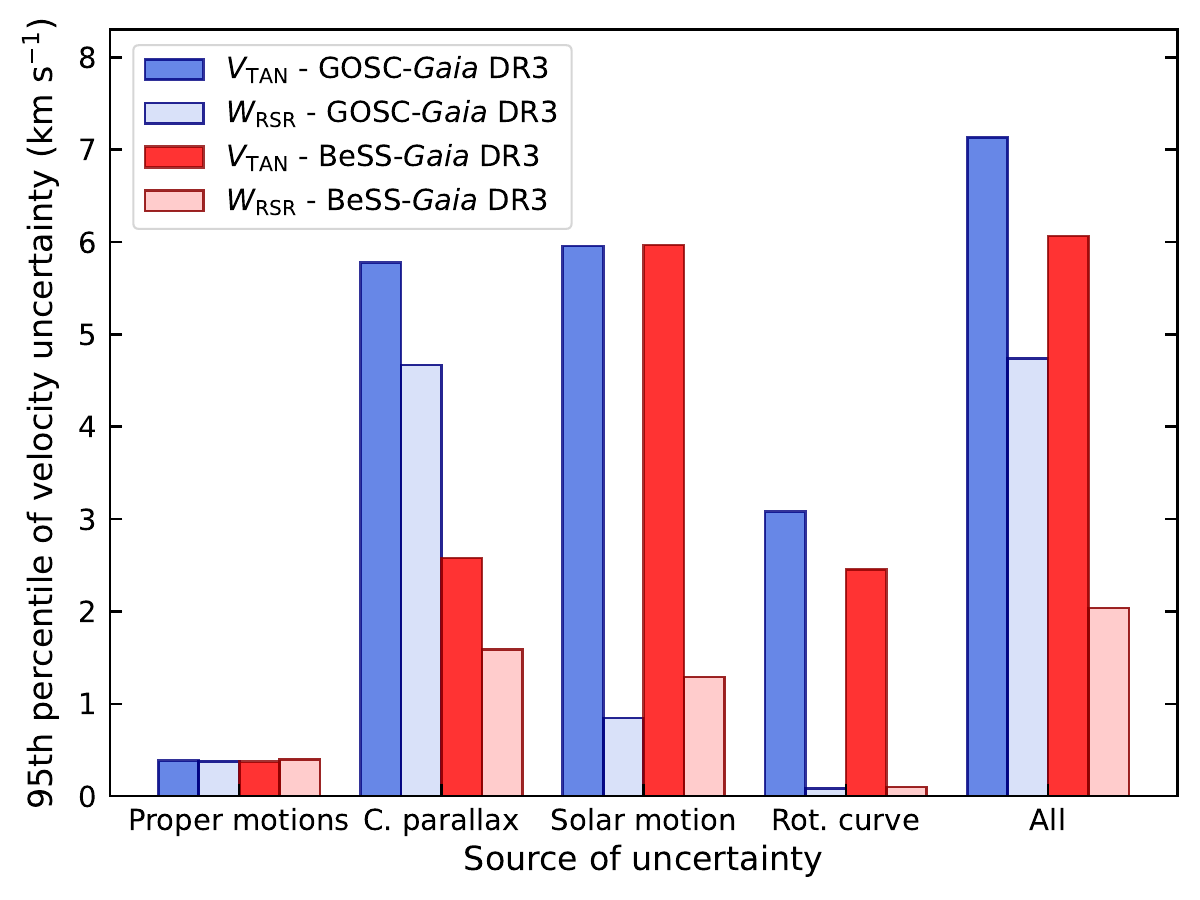}
    \caption{Ninety-fifth percentile of the $V_{\text{TAN}}$ and $W_{\text{RSR}}$ velocity uncertainties for different sources of uncertainty, and considering all the uncertainties together, for GOSC-\textit{Gaia}~DR3 stars in blue and BeSS-\textit{Gaia}~DR3 stars in red.}
    \label{Fig:Uncertainties}
   \end{figure}

The uncertainties in $V_{\text{TAN}}$ for the GOSC-\textit{Gaia}~DR3 catalog range between 1.5 and 35.8~km~s$^{-1}$, while 95\% of the stars have uncertainties smaller than 7.1~km~s$^{-1}$. For $W_{\text{RSR}}$, these numbers are 0.7, 24.9, and 4.7~km~s$^{-1}$. The uncertainties in $V_{\text{TAN}}$ for the BeSS-\textit{Gaia}~DR3 catalog range between 1.3 and 17.1~km~s$^{-1}$, while 95\% of the stars have uncertainties below 6.1~km~s$^{-1}$. For $W_{\text{RSR}}$, these numbers are 0.7, 14.0, and 2.0~km~s$^{-1}$. We show in Fig.~\ref{Fig:Uncertainties} the 95th percentile of the $V_{\text{TAN}}$ and $W_{\text{RSR}}$ velocity uncertainties for different source of uncertainty and considering all the uncertainties together (which correspond to the numbers listed above). The different sources of uncertainty considered are the proper motions $\left(\mu_{\alpha^*}\text{ and } \mu_{\delta}\right)$, the corrected parallax $\left(\varpi_{\text{c}}\right)$, the solar motion $\left(U_{\sun}, V_{\sun}, W_{\sun}\right)$, and the uncertainties associated with the Galactic rotation curve $\left(R_\sun, \Theta_\sun, d\Theta/dR\right)$. Uncertainties in position are not shown because they were negligible and had no influence on the velocity uncertainties. Figure~\ref{Fig:Uncertainties} shows that the uncertainties in proper motion produce velocity uncertainties around 0.4~km~s$^{-1}$ in all cases. The velocity uncertainties for $V_{\text{TAN}}$ are dominated by the uncertainties in the solar motion (particularly due to $V_{\sun}=13.6\pm6.8$~km~s$^{-1}$) in the case of BeSS-\textit{Gaia}~DR3, while an additional contribution from the uncertainty in corrected parallax takes place for GOSC-\textit{Gaia}~DR3. The contribution from the uncertainties related to the Galactic rotation curve are only about 2--3~km~s$^{-1}$. The velocity uncertainties for $W_{\text{RSR}}$ are dominated by the uncertainties in the corrected parallax owing to the small uncertainty of the solar motion perpendicular to the Galactic plane ($W_{\sun}=7.6\pm1.0$~km~s$^{-1}$) and to the lack of influence of uncertainties of the Galactic rotation curve in the velocity perpendicular to the Galactic plane. Considering all the uncertainties together, the $V_{\text{TAN}}$ uncertainties are larger than the corresponding $W_{\text{RSR}}$ uncertainties mainly because of the uncertainties related to the corrected parallax, the solar motion, and the Galactic rotation curve. We also note that the GOSC-\textit{Gaia}~DR3 uncertainties are generally larger than those of the BeSS-\textit{Gaia}~DR3. The reason is that O stars lie at larger distances on average than Be stars and that the applied correction to their parallaxes and corresponding uncertainties are also larger (because of their magnitudes and colors according to \citealt{MA2022}).

\subsection{Field stars versus runaway stars} \label{Sec:FSvsRun}

Runaway stars have historically been classified as stars that move with a significant peculiar velocity, either in two or three dimensions, that is, also above a certain threshold. This threshold was initially placed at 40~km~s$^{-1}$ \citep{Blaauw1961}. However, it was reduced to 30~km~s$^{-1}$ by \cite{Gies1987} based on a radial velocity component that is 3 times the standard deviation of the residual peculiar radial velocity for the bulk of O stars in the sample. This threshold has been further reduced to $\sim$25--28~km~s$^{-1}$ by more recent works \citep{Tetzlaff2011, Kobulnicky2022}. This reduction is based on the results of \cite{Tetzlaff2011}, who found that the peculiar 2D tangential velocity had two components described by two Maxwellian distributions: a low- and high-velocity component, which intersect at 20~km~s$^{-1}$. It is therefore sufficient to place the threshold at around $\sim$25~km~s$^{-1}$  to separate the stars of the high-velocity group. In addition, \cite{Boubert2018} found a median runaway velocity for their Be stars of 19.6~km~s$^{-1}$, which is even lower than the $\sim$25~km~s$^{-1}$ threshold. This would imply a significant loss of their runaways.  \cite{Stone1991} already noted that it would be better to search for runaways with a method independent of a runaway velocity threshold, and they suggested to do this by fitting a bimodal distribution function to the velocity histograms. Our samples are not large enough to properly trace the distribution for runaway stars. However, we used Gaussian functions to characterize the velocity distributions of field stars, which allowed us to classify the stars with significantly higher velocities as runaways without the need of a velocity threshold.

In this work, we define a runaway star as an star with a high peculiar 2D velocity $\left(V_{\text{TAN}},W_{\text{RSR}}\right)$ with respect to the mean Galactic rotation at a 3-sigma confidence level (but see below). This was achieved through the following parameter:
\begin{equation}
    \label{Eq:Criteria}
        E =\sqrt{ \left(\frac{V_{\text{TAN}} - \mu_{V_{\text{TAN}}}^{\text{GF}}}{3\Bar{\sigma}_{V_{\text{TAN}}}}\right)^2 + \left(\frac{W_{\text{RSR}} - \mu_{W_{\text{RSR}}}^{\text{GF}}}{3\Bar{\sigma}_{W_{\text{RSR}}}}\right)^2
        },
\end{equation}
where $\mu_{V_{\text{TAN}}}^{\text{GF}}$ and $\mu_{W_{\text{RSR}}}^{\text{GF}}$ are the means of the velocity distributions obtained from the Gaussian fits, and $\Bar{\sigma}_{V_{\text{TAN}}}$ and $\Bar{\sigma}_{W_{\text{RSR}}}$ are velocity uncertainties computed using the following equations: $\Bar{\sigma}_{V_\text{TAN}} = \sqrt{{\sigma_{V_\text{TAN}}^{\text{GF}}}^2 + \sigma_{V_{\text{TAN},*}}^2}$ and $\Bar{\sigma}_{W_\text{RSR}} = \sqrt{{\sigma_{W_\text{RSR}}^{\text{GF}}}^2 + \sigma_{W_{\text{RSR},*}}^2}$. These last equations take the standard deviations of the velocity distributions obtained from the Gaussian fits $\sigma_{V_\text{TAN}}^{\text{GF}}$ and $\sigma_{W_\text{RSR}}^{\text{GF}}$ into account, as well as the individual uncertainties of the velocities of each star $\sigma_{V_{\text{TAN},*}}$ and $\sigma_{W_{\text{RSR},*}}$. We note that in all this procedure, we evaluated how significant the $V_\text{TAN}$ and $W_{\text{RSR}}$ velocities of the stars are with respect the mean velocities defined by the stars themselves (not strictly with respect to the Galactic rotation curve). The stars with $E>1$ were classified as runaways. However, the means and standard deviations obtained from the Gaussian fits, particularly the latter, are affected by the presence of runaway stars. Therefore, we needed to proceed in an iterative way using a 3-sigma clipping algorithm as follows:

\begin{enumerate}
    \item Using Eq.~(\ref{Eq:Criteria}), we identified all the stars with $E>1.$
    \item We produced histograms without these stars and performed new Gaussian fits in both velocities.
    \item Using Eq.~(\ref{Eq:Criteria}) with updated values of 
    $\mu_{V_{\text{TAN}}}^{\text{GF}}$, $\mu_{W_{\text{RSR}}}^{\text{GF}}$, $\Bar{\sigma}_{V_\text{TAN}}$ , and $\Bar{\sigma}_{W_\text{RSR}}$ , we identified stars with $E>1.$
    \item When we found more stars that fulfilled $E>1$, we returned to point 2, otherwise, we stopped.
\end{enumerate}

After this clipping process, the means and standard deviations of the Gaussian fits represent the distribution of the field stars better. At this point, stars with $E>1$ were classified as runaways, and stars with $E<1$ were classified as field stars.

\section{Search for runaways: Results}\label{sec:results}

In this section, we present our results for the velocity and spatial distributions of the field and the runaway stars. We also present the results of an additional search for runaways using our method for specific spectral type bins.

\subsection{Velocity and spatial distributions}\label{Sec:Runaways_Found}

We show in Table~\ref{Tab:FitsAfterClippingField} the means and standard deviations for the field stars after clipping of the Gaussian fits applied to the $V_\text{TAN}$ and $W_{\text{RSR}}$ velocity distributions (using a bin size of 2~km~s$^{-1}$), and of the $Z$ and $b$ distributions for the complete and nearby (d $<2$~kpc) GOSC-\textit{Gaia}~DR3 and BeSS-\textit{Gaia}~DR3 catalogs. We note that $\sigma_{V_{\text{TAN}}}^{\text{GF}}$ is larger than $\sigma_{W_{\text{RSR}}}^{\text{GF}}$ for both catalogs, with a higher value of $\sigma_{V_{\text{TAN}}}^{\text{GF}}$ for the BeSS- stars than for those from GOSC-\textit{Gaia}~DR3. In all cases, the mean values are closer to zero than the bin size. The mean $Z$ and $b$ values are close to zero with some dispersion.

\renewcommand{\arraystretch}{1.1}
\begin{table*}
\caption{Means and standard deviations of different distributions for the field stars after clipping.}
\label{Tab:FitsAfterClippingField}
\centering
\begin{tabular}{l@{~~~~}c@{~~~~}cr@{~~$\pm$~~}lr@{~~$\pm$~~}lr@{~~$\pm$~~}lr@{~~$\pm$~~}l}
\hline \hline \vspace{-3mm}\\
Catalog  & Stars & Field Stars & 
$\mu_{V_{\text{TAN}}}^{\text{GF}}$ & $\sigma_{V_{\text{TAN}}}^{\text{GF}}$ &
$\mu_{W_{\text{RSR}}}^{\text{GF}}$ & $\sigma_{W_{\text{RSR}}}^{\text{GF}}$ &
$\mu_Z$ & $\sigma_Z$ &
$\mu_b$ & $\sigma_b$ \\
 & \# & \#
 & \multicolumn{2}{c}{(km~s$^{-1}$)} 
 & \multicolumn{2}{c}{(km~s$^{-1}$)}
 & \multicolumn{2}{c}{(kpc)} 
 & \multicolumn{2}{c}{($\degr$)} 
\\
\hline \vspace{-3mm}\\
GOSC-\textit{Gaia}~DR3              & ~~417 & ~~311 &     1.0 & 6.6   &  $-$1.0 & 5.3   &  $-$0.01 & 0.07  &  $-0.1$ & 2.0 \\
BeSS-\textit{Gaia}~DR3              &  1335 &  1266 &     1.6 & 9.3   &  $-$0.5 & 4.9   &  $-$0.01 & 0.11  &  $-1.0$ & 6.5 \\
\hline
GOSC-\textit{Gaia}~DR3 ($d<2$~kpc)  & ~~197 & ~~145 &  $-$0.1 & 5.0   &     1.4 & 4.2   &     0.02 & 0.04  &     0.6 & 2.1 \\
BeSS-\textit{Gaia}~DR3 ($d<2$~kpc)  & ~~831 & ~~793 &     0.6 & 9.7   &  $-$0.4 & 5.1   &  $-$0.02 & 0.10  &  $-1.5$ & 8.2 \\
\hline
\end{tabular}
\end{table*}

\renewcommand{\arraystretch}{1.1}
\begin{table*}
\caption{Means and standard deviations of different distributions for the runaway stars after clipping.}
\label{Tab:FitsAfterClippingRun}
\centering
\begin{tabular}{l@{~~~~}c@{~~~~}cr@{~~$\pm$~~}lr@{~~$\pm$~~}lr@{~~$\pm$~~}lr@{~~$\pm$~~}l}
\hline \hline \vspace{-3mm}\\
Catalog  & Stars & Runaway Stars &
$\mu_{V_{\text{TAN}}}$ & $\sigma_{V_{\text{TAN}}}$ &
$\mu_{W_{\text{RSR}}}$ & $\sigma_{W_{\text{RSR}}}$ &
$\mu_Z$ & $\sigma_Z$ &
$\mu_b$ & $\sigma_b$ \\
 & \# & \#
 & \multicolumn{2}{c}{(km~s$^{-1}$)} 
 & \multicolumn{2}{c}{(km~s$^{-1}$)}
 & \multicolumn{2}{c}{(kpc)} 
 & \multicolumn{2}{c}{($\degr$)} 
\\
\hline \vspace{-3mm}\\
GOSC-\textit{Gaia}~DR3              & ~~417 &  106 &  $-$3.1 & 39.7  &  1.1 & 38.7  &     0.01 & 0.19  &    0.2  & ~~4.3 \\
BeSS-\textit{Gaia}~DR3              &  1335 & ~~69 &  $-$7.7 & 34.6  &     1.0 & 22.3  &  $-$0.09 & 0.66  &  $-3.4$ &  21.2 \\
\hline
GOSC-\textit{Gaia}~DR3 ($d<2$~kpc)  & ~~197 & ~~52 &  $-$3.9 & 28.6  &  $-$0.6 & 32.3  &     0.00 & 0.11  &     $-$0.1 & ~~4.1 \\
BeSS-\textit{Gaia}~DR3 ($d<2$~kpc)  & ~~831 & ~~38 &  $-$5.7 & 33.0  &     1.8 & 22.9  &  $-0.06$ & 0.40  &  $-3.2$ &  22.1 \\
\hline 
\end{tabular}
\end{table*}

After the clipping process discussed in Sect.~\ref{Sec:FSvsRun}, we obtained 106 runaway stars for the
GOSC-\textit{Gaia}~DR3 catalog  (42 newly published as such here, to our knowledge), which is equivalent to 25.4\% of the catalog. For the BeSS-\textit{Gaia}~DR3 catalog, we obtained 69 runaway stars (47 newly published as such here), representing 5.2\% of the catalog. 
We note that the already known runaway Oe star \object{HD~57682} is also classified as runaway in the GOSC-\textit{Gaia}~DR3 catalog.
We show in Table~\ref{Tab:FitsAfterClippingRun} the means and standard deviations as in Table~\ref{Tab:FitsAfterClippingField}, but for the runaway stars and without using Gaussian fits for the velocity distributions. As expected, the velocity dispersions are larger for the runaways than for the field stars, with values in the range 20--40~km~s$^{-1}$. This agrees with the dispersions of 25--30~km~s$^{-1}$ in the literature \citep{Stone1991,Tetzlaff2011}. Finally, the dispersions in $Z$ and $b$ are larger than for the field stars (see below).

\begin{figure}
    \centering
   \includegraphics[width=\hsize]{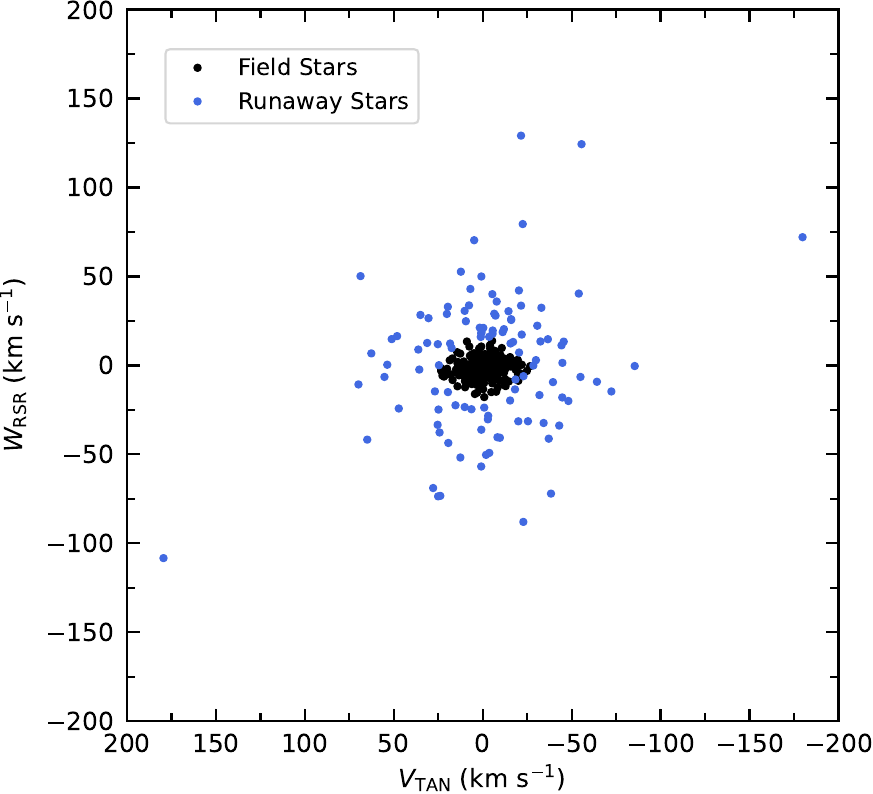}
    \caption{$W_{\text{RSR}}$ as a function of $V_{\text{TAN}}$ for the GOSC-\textit{Gaia}~DR3 stars. Field stars are depicted in black, and runaway stars are shown in blue.}
    \label{Fig:GOSC_2D}
\end{figure}

\begin{figure}
    \centering
    \includegraphics[width=\hsize]{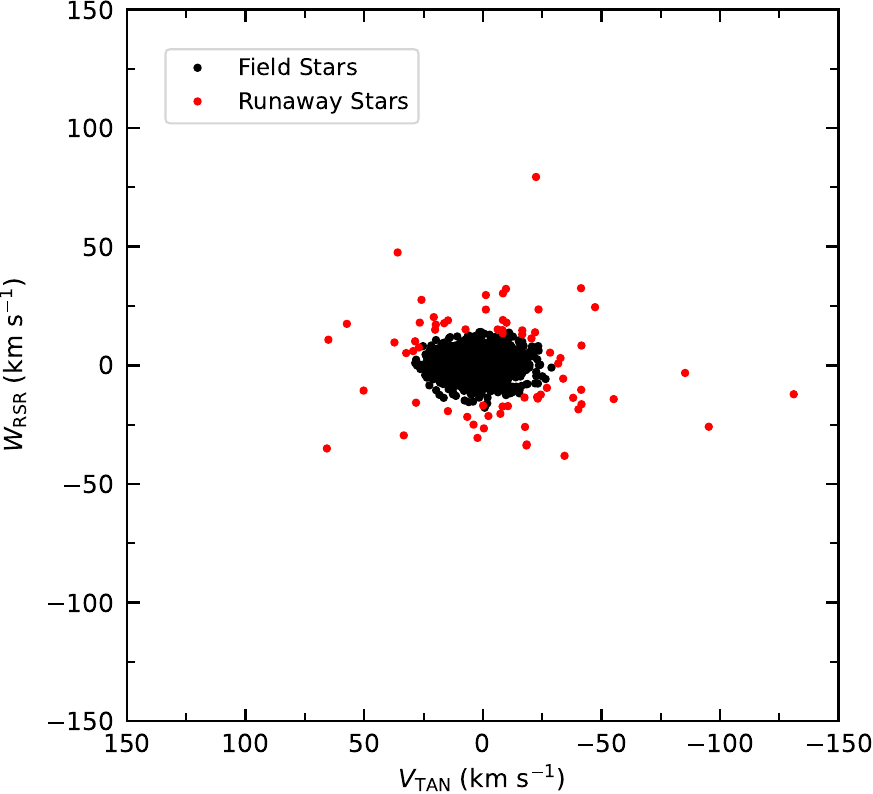}
    \caption{$W_{\text{RSR}}$ as a function of $V_{\text{TAN}}$ for the BeSS-\textit{Gaia}~DR3 stars. Field stars are depicted in black, and runaway stars are shown in red.}
    \label{Fig:BeSS_2D}
\end{figure}

The 2D $\left(V_{\text{TAN}},W_{\text{RSR}}\right)$ velocity distribution of the GOSC-\textit{Gaia}~DR3 stars is displayed in Fig.~\ref{Fig:GOSC_2D}. Runaway stars, which are indicated in blue, clearly have high velocities, whereas field stars, which are represented in black, have velocities about (0,0) and are approximately contained in a 2D ellipse (if all velocity uncertainties of the individual stars were the same, it would be a real ellipse). We note that many of the black circles for the field stars overlap because the number of stars with low velocities is high, as shown in Fig.~\ref{Fig:GOSC_VTAN_WRSR}. The 2D velocity distribution of the BeSS-\textit{Gaia}~DR3 stars is displayed in Fig.~\ref{Fig:BeSS_2D}. Runaway stars are indicated in red and field stars in black (the scale is different from Fig.~\ref{Fig:GOSC_2D}). The observed behavior is similar as for the GOSC-\textit{Gaia}~DR3 stars, although in this case, the runaway stars have slightly lower velocities. We note that the standard deviations of the velocity distributions of the field stars, $\sigma_{V_{\text{TAN}}}^{\text{GF}}$ and $\sigma_{W_{\text{RSR}}}^{\text{GF}}$, presented in Table~\ref{Tab:FitsAfterClippingField} are fundamental for determining which stars are runaways. Higher values of the standard deviations imply a lower number of detected runaways. For both catalogs, $\sigma_{W_{\text{RSR}}}^{\text{GF}}$ is smaller than $\sigma_{V_{\text{TAN}}}^{\text{GF}}$. In addition, as presented in Sect.~\ref{Sec:Uncertainties}, the $V_{\text{TAN}}$ uncertainties are larger than those of $W_{\text{RSR}}$. This is due to the contribution of the Galactic rotation curve uncertainty, which affects the $V_{\text{TAN}}$ component but not $W_{\text{RSR}}$. Both effects translate into more runaways that are found in the $W_\text{RSR}$ component, as shown in Figs.~\ref{Fig:GOSC_2D}~and~\ref{Fig:BeSS_2D}, where the smaller axis of the field star ellipse is obtained in $W_\text{RSR}$.

\begin{figure}[t]
    \centering
    \includegraphics[width=\hsize]{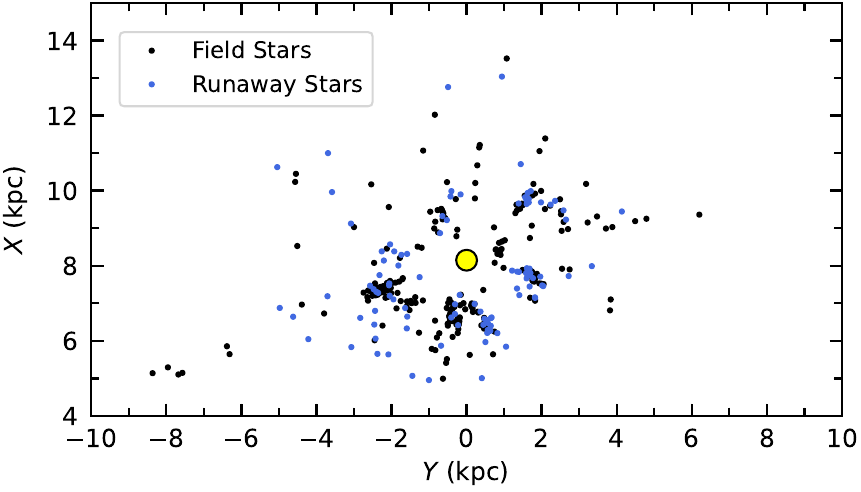}
    \caption{Galactocentric $XY$ coordinates for the GOSC-\textit{Gaia}~DR3 stars. Field stars are depicted in black, and runaway stars are shown in blue. The position of the Sun is marked with a yellow circle at $\left(X, Y\right) = \left(8.15,0\right)$~kpc.}
    \label{Fig:GOSC_XY}
\end{figure}

\begin{figure}[t]
    \centering
    \includegraphics[width=\hsize]{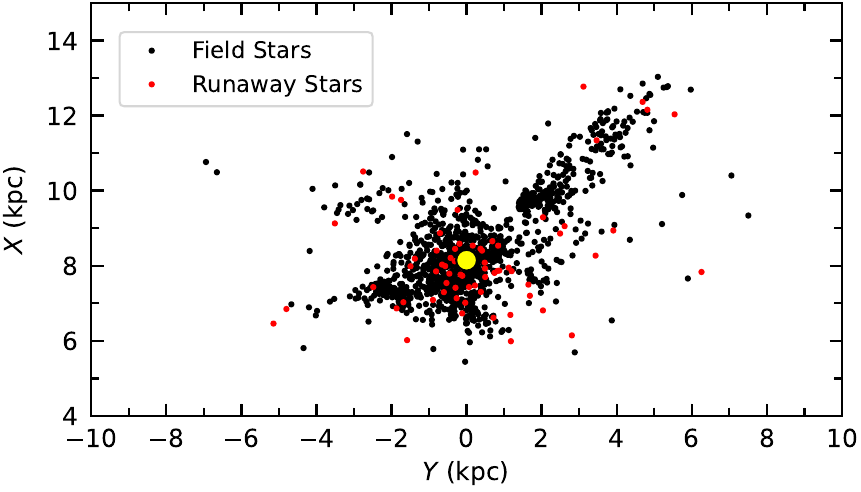}
    \caption{Same as Fig.~\ref{Fig:GOSC_XY} for the BeSS-\textit{Gaia}~DR3 stars. Field stars are depicted in black, and runaway stars are shown in red.}
    \label{Fig:BeSS_XY}
\end{figure}

The galactocentric $XY$ coordinates for the GOSC-\textit{Gaia}~DR3 and the BeSS-\textit{Gaia}~DR3 stars are shown in Figs.~\ref{Fig:GOSC_XY}~and~\ref{Fig:BeSS_XY}, respectively. As noted when presenting Fig.~\ref{Fig:XYGal}, there is an absence of GOSC-\textit{Gaia}~DR3 stars close to the Sun due to the cut in $G$ magnitude. This cut also affects BeSS-\textit{Gaia}~DR3 stars, but the effect is not visible in the figures because of the many fainter objects. Therefore, there is a significant number of Be runaway stars close to the Sun. The figures do not show any privileged direction from the Sun with a significantly larger or smaller number of runaway stars. There is also no significant difference as a function of galactocentric distance.

Figures of the galactocentric $XZ$ and $YZ$ coordinates for both catalogs are presented in Appendix~\ref{Sec:App_ZPlanes}. The mean and standard deviation in the $Z$ coordinate for the field and the runaway stars are presented in Tables~\ref{Tab:FitsAfterClippingField}~and~\ref{Tab:FitsAfterClippingRun}, respectively. For the GOSC-\textit{Gaia}~DR3 catalog, the dispersion in $Z$ for the runaways is about three times larger than for the field stars. For the BeSS-\textit{Gaia}~DR3 catalog, this dispersion is  about six times larger for the runaway stars. When limiting the catalog to $d<2$~kpc, we typically obtained lower values for both field and runaway stars.

\begin{figure*}
    \centering
    \includegraphics[width=0.78\hsize]{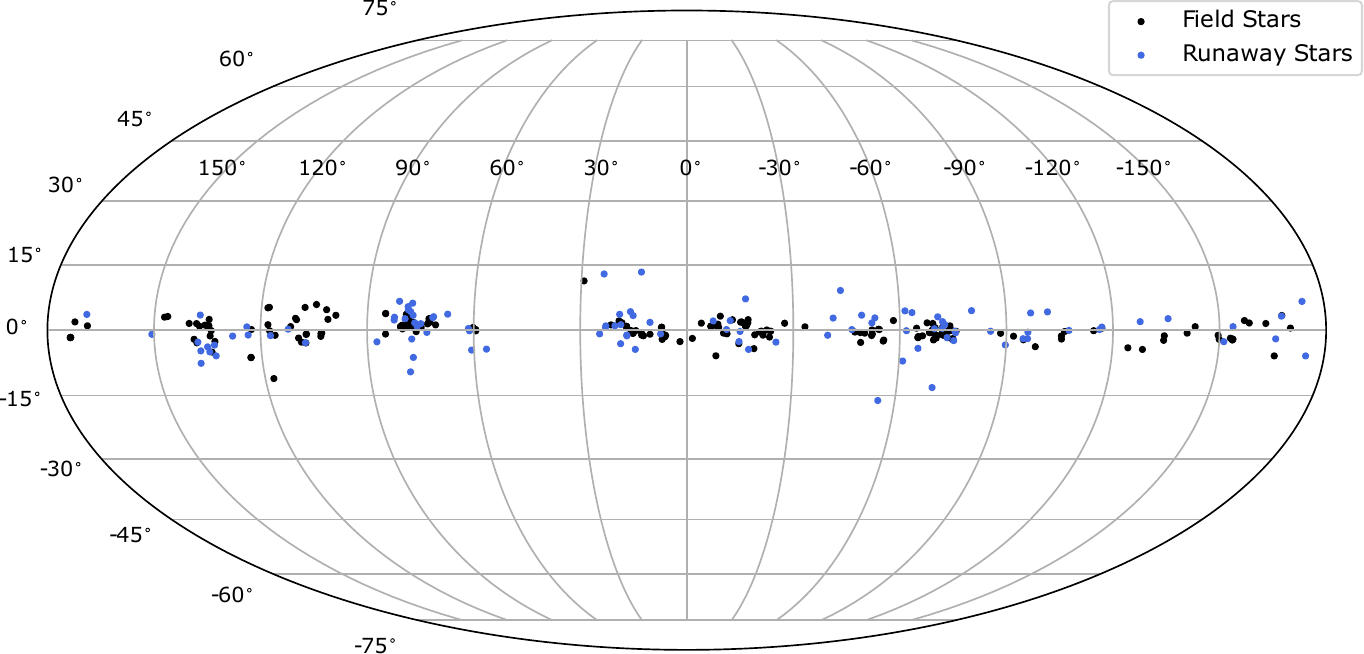}
    \caption{Galactic $\left(l,b\right)$ coordinates for GOSC-\textit{Gaia}~DR3 stars. Field stars are depicted in black, and runaway stars are shown in blue.}
    \label{Fig:GOSC_Aitoff}
\end{figure*}

\begin{figure*}
    \centering
    \includegraphics[width=0.78\hsize]{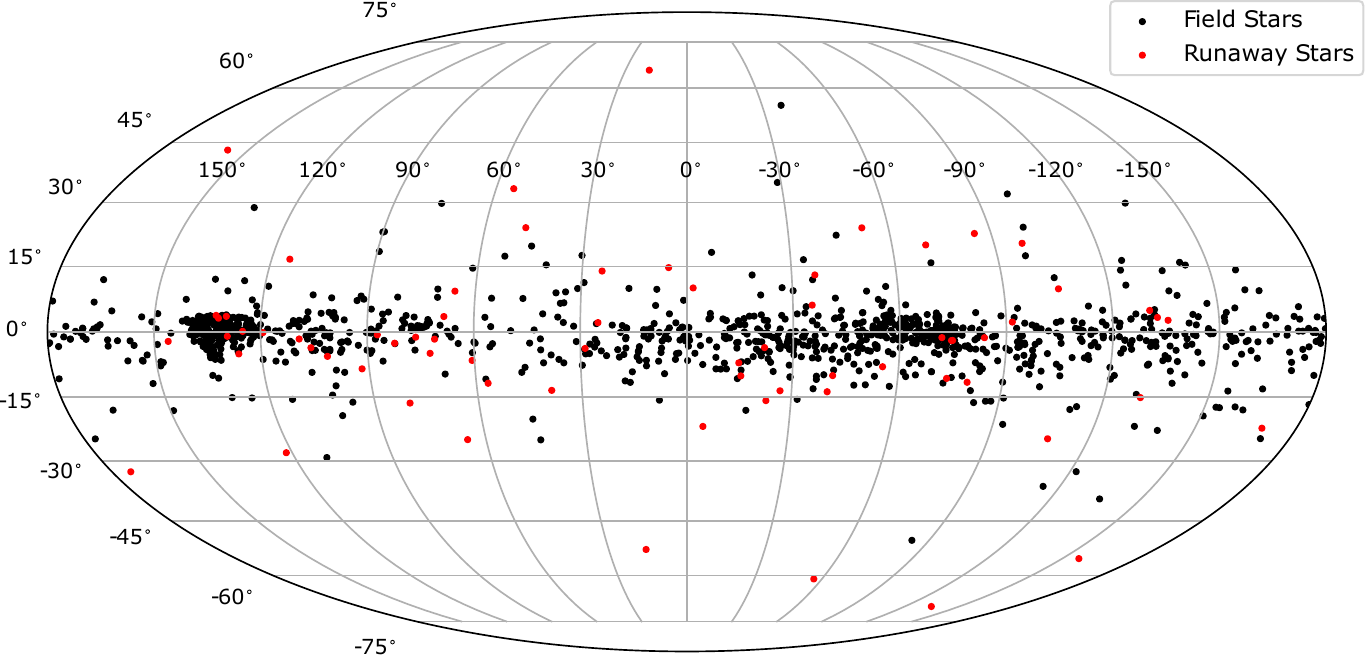}
    \caption{Galactic $\left(l,b\right)$ coordinates for BeSS-\textit{Gaia}~DR3 stars. Field stars are depicted in black, and runaway stars are shown in red.}
    \label{Fig:BeSS_Aitoff}
\end{figure*}

The projections in the $\left(l, b\right)$ Galactic coordinates for the GOSC- and the BeSS-\textit{Gaia}~DR3 stars are displayed in Figs.~\ref{Fig:GOSC_Aitoff}~and~\ref{Fig:BeSS_Aitoff}, respectively. In both cases, the stars are clustered around Galactic latitude $b=0\degr$ with some dispersion. The mean and standard deviations in $b$ for the field and the runaway stars are presented in the last column of Tables~\ref{Tab:FitsAfterClippingField}~and~\ref{Tab:FitsAfterClippingRun}, respectively. For the GOSC-\textit{Gaia}~DR3 catalog, the dispersion in $b$ for the runaways is about twice larger than for the field stars. For the BeSS-\textit{Gaia}~DR3 catalog, this dispersion is about three times larger for the runaway stars. The standard deviation for the BeSS-\textit{Gaia}~DR3 stars is larger than for the GOSC-\textit{Gaia}~DR3, indicating that the last stars are more clustered and closer to the Galactic plane. When limiting the catalog to $d<2$~kpc, we obtain similar values in all cases.

In Tables~\ref{Tab:GOSC_Runaways}~and~\ref{Tab:BeSS_Runaways} we show ten rows of data corresponding to the runaways with the higher $E$ values, in decreasing order, of the GOSC and BeSS-\textit{Gaia}~DR3 catalogs, respectively. They include their names and identifications, coordinates, distances, magnitudes, spectral types, velocities, $E$ values, a column indicating the reference in the cases that they had already been reported as runaways, and a last column with additional information. In these tables, we include the 2D peculiar velocity, which is computed as $V_\text{PEC}^\text{2D} = \sqrt{V_\text{TAN}^2 + W_\text{RSR}^2}$. For the GOSC-\textit{Gaia}~DR3 runaway stars, it ranges from $16$ to $210$~km~s$^{-1}$ and has a median (mean) of $\sim$40 ($\sim$46)~km~s$^{-1}$. For the BeSS-\textit{Gaia}~DR3 runaway stars, $V_\text{PEC}^\text{2D}$ ranges from $16$ to $132$~km~s$^{-1}$ and has a median (mean) of $\sim$30 (37)~km~s$^{-1}$. The complete versions of these tables, including all classified runaways with additional parameters and precision, are available at the CDS.

Finally, we note that the bin size of the $V_\text{TAN}$ and $W_\text{RSR}$ histograms is relevant when we apply Eq.~(\ref{Eq:Criteria}) to classify the runaway stars. In this work, we used a bin size of 2~km~s$^{-1}$. However, we tried with a bin size of 3~km~s$^{-1}$ and the results did not change in the classification of the GOSC-\textit{Gaia}~DR3 runaways. For the BeSS-\textit{Gaia}~DR3 catalog, we only lost two runaways. The mean and standard deviations obtained from the Gaussian fits did not change significantly, except for $\sigma_{V_{\text{TAN}}}^{\text{GF}}$ of the BeSS-\textit{Gaia}~DR3 catalog, which changed from 9.3 to 9.5~km~s$^{-1}$.

\renewcommand{\arraystretch}{1.1}
\begin{table*}
\centering
\tiny
\caption{Data of the 10 GOSC-\textit{Gaia}~DR3 runaway stars with the higher values of $E$ in decreasing order.}
\label{Tab:GOSC_Runaways}
\resizebox{\textwidth}{!}{\begin{tabular}{l@{~~~}c@{~~~}r@{~}c@{~}l@{~~~}r@{~}c@{~}l@{~~~}c@{~~~}c@{~~~}c@{~~~}c@{~~~}c@{~~~}r@{$~\pm$~}l@{~~~}r@{$~\pm~$}l@{~~~}r@{$~\pm~$}l@{~~~}c@{~~~}c@{~~~}c}
\hline\hline \vspace{-2mm}\\
GOSC Name & \textit{Gaia} DR3 id & \multicolumn{3}{c}{RA} & \multicolumn{3}{c}{DEC} & $l$ & $b$ & $d$ & $G$ & S.T. & \multicolumn{2}{c}{$V_\text{TAN}$} & \multicolumn{2}{c}{$W_\text{RSR}$} & \multicolumn{2}{c}{$V_\text{PEC}^\text{2D}$} & $E$ & Run. Ref. & Comments \\
& & (hh & mm &ss.ss) & ($\degr$ & $\arcmin$ & $\arcsec.\arcsec$) & ($\degr$) & ($\degr$) & (kpc) & & & \multicolumn{2}{c}{(km~s$^{-1}$)} & \multicolumn{2}{c}{(km~s$^{-1}$)} & \multicolumn{2}{c}{(km~s$^{-1}$)} & & \\
\hline \vspace{-2mm}\\
  V479 Sct       & 4104196427943626624 & 18 & 26 & 15.06 & $-$14 & 50 & 54.4 & ~~16.9 & $-$1.3 & 1.94 &  10.80 & ON6  &  $-$22.8 & 5.9  & $-$88.0 & 4.5  & 90.9  & 4.9     & 4.28 & 1 & h,g\\
  CPD $-$34 2135 & 5546501254035205376 & 08 & 13 & 35.36 & $-$34 & 28 & 43.9 &  252.4 & $-$0.1 & 3.23 & ~~9.18 & O7.5 &     68.8 & 4.8  & 50.1    & 3.6  & 85.1  & 5.6     & 3.85 & 2 & ---\\
  BD $+$60 134   & 427457895747434880  & 00 & 56 & 14.21 &    61 & 45 & 36.9 &  123.5 & $-$1.1 & 2.69 &  10.40 & O5.5 &     25.1 & 4.0  & $-$73.7 & 4.3  & 77.9  & 4.8     & 3.70 & 3 & f\\
  HD 104 565     & 6072058878595295488 & 12 & 02 & 27.77 & $-$58 & 14 & 34.3 &  296.5 &  ~~4.0 & 4.72 & ~~9.07 & OC9.7& $-$179.7 & 24.0 & 72.0    & 8.3  & 193.7 & 25.2    & 3.47 & 2 & ---\\
  HD 155 913     & 5953699131931631232 & 17 & 16 & 26.32 & $-$42 & 40 & 04.0 &  345.3 & $-$2.6 & 1.21 & ~~8.18 & O4.5 &      4.9 & 5.8  & 70.3    & 4.5  & 70.6  & 4.5     & 3.43 & 2 & f\\
  HD 75 222      & 5625488726258364544 & 08 & 47 & 25.13 & $-$36 & 45 & 02.5 &  258.3 &  ~~4.2 & 2.09 & ~~7.31 & O9.7 &  $-$53.9 & 6.6  & 40.3    & 2.8  & 67.4  & 6.9     & 3.03 & 2 & ---\\
  ALS 12 688     & 2003815312630659584 & 22 & 55 & 44.94 &    56 & 28 & 36.6 &  107.4 & $-$2.9 & 4.34 &  10.39 & O5.5 &     27.9 & 3.6  & $-$69.0 & 6.8  & 74.5  & 6.8     & 2.90 & 3 & ---\\
  BD $-$14 5040  & 4104201586232296960 & 18 & 25 & 38.91 & $-$14 & 45 & 05.8 & ~~16.9 & $-$1.1 & 1.63 &  10.02 & O5.5 &   $-$1.8 & 5.8  & $-$50.3 & 2.2  & 50.4  & 2.3     & 2.87 & 2 & ---\\
  HD 41 997      & 3345950879898371712 & 06 & 08 & 55.82 &    15 & 42 & 18.0 &  194.2 & $-$2.0 & 1.75 & ~~8.30 & O7.5 &     65.0 & 8.6  & $-$41.8 & 4.3  & 77.3  & 9.1     & 2.79 & 2 & f\\
  Y Cyg          & 1869256701670871168 & 20 & 52 & 03.58 &    34 & 39 & 27.2 & ~~77.3 & $-$6.2 & 1.39 & ~~7.26 & O9.5 &  $-$38.3 & 4.1	& $-$72.1 & 9.3  & 82.1  & 10.0    & 2.78 & 2 & ---\\
\hline
\end{tabular}}
\tablefoot{Run. Ref. (Runaway Reference) column provides a reference in case the star had already been reported as runaway: (1)~\citet{Ribo2002}; (2) \citet{MA2018}; (3) \citet{Kobulnicky2022}. Comments column contains additional information: `f' for fast rotator stars \citep{Britavskiy2023}; `b' for binary hints \citep{Brandt2021}, `h' for high-mass X-ray binaries \citep{Fortin2023}; `g' for gamma-ray binaries \citep{Bordas2023}. The complete version of this table, including the 106 classified runaways with additional parameters and precision, is available at the CDS}.
\end{table*}

\renewcommand{\arraystretch}{1.1}
\begin{table*}
\centering
\tiny
\caption{Data of the 10 BeSS-\textit{Gaia}~DR3 runaway stars with the higher values of $E$ in decreasing order.} 
\label{Tab:BeSS_Runaways}
\centering
\resizebox{\textwidth}{!}{\begin{tabular}{l@{~~~}c@{~~~}r@{~}c@{~}l@{~~~}r@{~}c@{~}l@{~~~}c@{~~~}c@{~~~}c@{~~~}c@{~~~}c@{~~~}r@{$~\pm$~}l@{~~~}r@{$~\pm~$}l@{~~~}r@{$~\pm~$}l@{~~~}c@{~~~}c@{~~~}c}
\hline\hline \vspace{-2mm}\\
BeSS Name & \textit{Gaia} DR3 id & \multicolumn{3}{c}{RA} & \multicolumn{3}{c}{DEC} & $l$ & $b$ & $d$ & $G$ & S.T. & \multicolumn{2}{c}{$V_\text{TAN}$} & \multicolumn{2}{c}{$W_\text{RSR}$} & \multicolumn{2}{c}{$V_\text{PEC}^\text{2D}$} & $E$ & Run. Ref. & Comments \\
& & (hh & mm &ss.ss) & ($\degr$ & $\arcmin$ & $\arcsec.\arcsec$) & ($\degr$) & ($\degr$) & (kpc) & & & \multicolumn{2}{c}{(km~s$^{-1}$)} & \multicolumn{2}{c}{(km~s$^{-1}$)} & \multicolumn{2}{c}{(km~s$^{-1}$)} & & \\
\hline \vspace{-2mm}\\
  CD$-$29 6963 & 5641493526747324416 & 08 & 56 & 48.24 & $-$30 & 11 & 01.8 &  254.4 &  ~~~~9.8 & 3.70 & ~~9.15  & Be   & 35.9     & 3.3  & 47.6    & 3.8 & 60.2 & 3.8   & 2.85 & --- & ---\\
  HD 181409    & 2049168552364523520 & 19 & 19 & 03.72 &    33 & 23 & 19.2 & ~~65.8 &  ~~~~9.3 & 0.55 & ~~6.55  & B2e  & $-$85.3  & 5.8  & $-$3.2  & 1.3 & 85.3  & 5.8  & 2.64 & 1   & ---\\
  HD 114200    & 5843657293789559296 & 13 & 10 & 52.71 & $-$70 & 48 & 31.1 &  304.5 & ~~$-$8.0 & 2.29 & ~~8.35  & B1e  & 65.8     & 7.0  & $-$35.1 & 4.3 & 74.7  & 8.0  & 2.55 & --- & ---\\
  BD$-$21 1449 & 2938176961013250432 & 06 & 25 & 16.94 & $-$21 & 20 & 04.3 &  229.5 &  $-$15.1 & 2.70 &  10.43  & Be   & $-$34.4  & 5.6  & $-$38.2 & 3.1 & 52.3  & 5.7  & 2.44 & --- & ---\\
  EM* StHA 143 & 1333493685058234880 & 17 & 14 & 25.38 &    31 & 35 & 00.4 & ~~54.5 &   ~~33.4 & 4.12 &  11.70  & Be   & $-$131.1 & 17.1 & $-$12.2 & 5.6 & 131.7 & 17.5 & 2.33 & --- & ---\\
  HD 77147     & 5297543543432635008 & 08 & 57 & 23.29 & $-$63 & 29 & 49.1 &  280.1 &  $-$11.6 & 0.68 & ~~8.42  & B8   & 65.1     & 2.3  &    10.8 & 1.4 & 66.1  & 2.2  & 2.33 & --- & ---\\
  HD 127617    & 1239086696118147712 & 14 & 31 & 59.56 &    18 & 46 & 00.5 & ~~18.4 &   ~~65.4 & 1.71 & ~~8.71  & B5e  & $-$95.2  & 15.5 & $-$25.9 & 3.4 & 105.7 & 16.2 & 2.28 &  2  & b\\
  BD$-$11 2043 & 3033914530822194688 & 07 & 39 & 42.24 & $-$12 & 15 & 33.3 &  229.4 &  ~~~~4.9 & 3.64 & ~~9.43  & Be   & $-$41.4  & 9.9  &    32.5 & 3.2 & 52.7  & 9.6  & 2.16 & --- & ---\\
  HD 159489    & 5955254318081779072 & 17 & 37 & 13.83 & $-$45 & 09 & 26.7 &  345.3 & ~~$-$7.1 & 1.05 & ~~8.15  & B1e  & $-$9.7   & 6.2  &    32.2 & 2.2 & 33.9  & 3.2  & 2.07 & --- & ---\\
  HD 43789     & 5567419359659568896 & 06 & 15 & 56.55 & $-$44 & 37 & 10.6 &  252.2 &  $-$24.7 & 0.92 & ~~8.51  & B6.5 & $-$18.4  & 2.6  & $-$33.8 & 2.9 & 41.0  & 3.6  & 2.07 & --- & ---\\
\hline
\end{tabular}}
\tablefoot{Run. Ref. column provides a reference in case the star had already been reported as runaway: (1)~\citet{Hoogerwerf2001}; (2) \citet{Boubert2018}. Comments column contains additional information: `f' for fast rotator stars \citep{Britavskiy2023}; `b' for binary hints \citep{Brandt2021}, `h' for high-mass X-ray binaries \citep{Fortin2023}; `g' for gamma-ray binaries \citep{Bordas2023}. The complete version of this table, including the 69 classified runaways with additional parameters and precision, is available at the CDS.}
\end{table*}

\subsection{Runaways as a function of spectral type} \label{sec:SpectralTypeRunawaysResults}

\begin{figure}[t!]
    \centering
    \includegraphics[width=0.9\hsize]{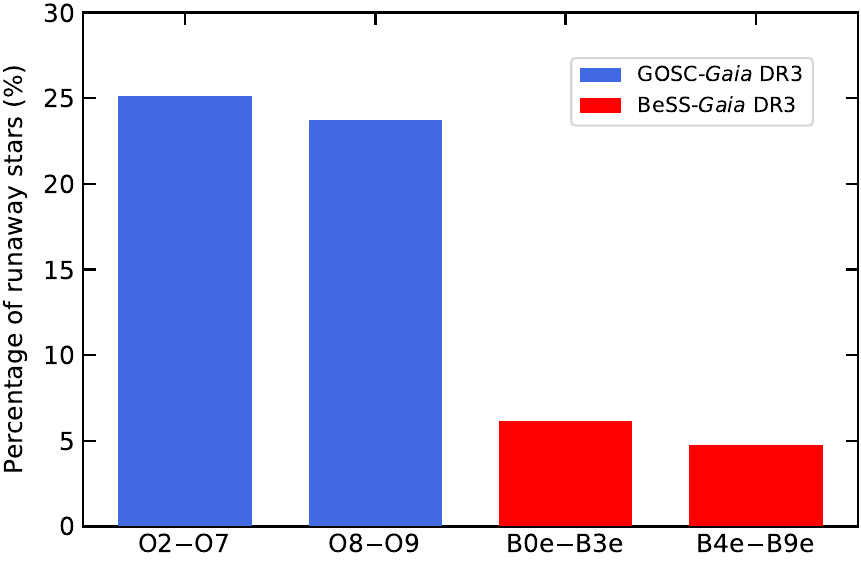}
    \caption{Percentage of runaway stars as a function of spectral type. Blue and red correspond to the GOSC-\textit{Gaia}~DR3 and BeSS-\textit{Gaia}~DR3 stars, respectively.}
      \label{Fig:Runaways_ST}
\end{figure}

We performed an additional search for runaways using our method for specific spectral type bins. We first removed stars from the GOSC-\textit{Gaia}~DR3 and BeSS-\textit{Gaia}~DR3 catalogs with an uncertain or variable spectral classification and binary systems with two spectral classifications. Next, we divided our samples into different spectral type bins. We tried to choose the bins as homogeneously as possible, considering that the stars are not homogeneously distributed in spectral type and that the GOSC and BeSS samples are different in size. We decided to use only two bins for each catalog to obtain significantly large samples (plus an additional bin in the intersection). To obtain the number of runaway stars, we applied Eq.~(\ref{Eq:Criteria}) for each bin independently. The results in number and percentage are shown in Table~\ref{Tab:SpectralType}. These results are similar to those obtained with the full samples although there is a trend of a decreasing percentage of runaway stars at later spectral types, as shown in Fig.~\ref{Fig:Runaways_ST}.
Because the percentage drops very significantly from the O to the Be stars, we conducted the search for an intermediate bin of spectral type O8--B1e composed of about 200 GOSC and 200 BeSS stars. The obtained results are shown in the last row of Table~\ref{Tab:SpectralType}. They reveal a percentage between those found previously.

\section{Discussion} \label{Sec:Discussion}

This section is organized into five subsections. First, we discuss the velocity dispersion of the field stars. Second, we examine the spatial distribution of the runaway stars. Third, we compare our results with previous works and discuss the implications of our findings. Fourth, we investigate the percentage of runaways as a function of spectral type and its relation to the origin of the runaway stars. Finally, we place our findings in the context of high-mass X-ray binaries and gamma-ray binaries.

\subsection{Velocity dispersion of field stars}\label{Sec:Discussion_Velocities}

The kinematics of OB field stars is related to that of the gas of the Milky Way disk. Because these stars are young, they should approximately follow the motion of the gas from which they recently formed \citep{Gontcharov2012,GaiaEDR32022_Disk}. In this context, we compared our velocity dispersions with those of two works. On the one hand, \cite{GaiaEDR32022_Disk} presented an average azimuthal and vertical dispersions of 9.4 and 6.2~km~s$^{-1}$ , respectively, for a sample of OB stars. Our dispersions of $\sim$7--9~km~s$^{-1}$ for $V_\text{TAN}$ and $\sim$5~km~s$^{-1}$ for $W_\text{RSR}$ (see Table~\ref{Tab:FitsAfterClippingField}) are slightly smaller. This may be because here we present the dispersions of a \textit{\textup{clean}} sample of stars without runaway stars, which would widen the distributions. We note that although the azimuthal velocity is not the same as our tangential velocity, their dispersions are expected to be comparable (see Eq.~(\ref{Eq:NewSystem})). On the other hand, \cite{Marasco2017} obtained a velocity dispersion of the gas of about 4--9~km~s$^{-1}$, which agrees very well with both \cite{GaiaEDR32022_Disk} and our results. 

As shown in Table~\ref{Tab:FitsAfterClippingField}, the dispersion of $W_\text{RSR}$ is around 5~km~s$^{-1}$ for both catalogs, while the dispersion in $V_\text{TAN}$ increases to 6.6~km~s$^{-1}$ for GOSC-\textit{Gaia}~DR3 stars and up to 9.3~km~s$^{-1}$ for BeSS-\textit{Gaia}~DR3 stars. Therefore, there is a velocity dispersion increase in the Galactic plane. This can be due to a combined effect of parallax (distance) uncertainties and the application of the Galactic rotation curve, which induces errors for distant stars in the tangential component of the velocity, but not in the vertical component. To study this possibility, we performed the same study, but restricted the distance to 2~kpc. The results, also shown in Table~\ref{Tab:FitsAfterClippingField}, reveal that the dispersion of $V_\text{TAN}$ for GOSC-\textit{Gaia}~DR3 stars decreases to 5.0~km~s$^{-1}$ and is thus very similar to the different dispersions obtained in $W_\text{RSR}$. It is therefore clear that the use of the Galactic rotation curve for stars with larger distances (and with larger uncertainties) artificially enhances this dispersion. However, in the case of BeSS-\textit{Gaia}~DR3 stars, the dispersion of $V_\text{TAN}$ does not decrease and seems to be intrinsic. This can be understood because Be stars are older than O stars and are therefore more affected by Galactic velocity diffusion in the disk, producing a larger velocity dispersion (see, e.g., Fig.~10 of \cite{Antoja2021} for trends in the same sense). We conclude that the velocity dispersion at birth for young stars is $\sim$5~km~s$^{-1}$, while it increases to higher values with age due to Galactic velocity diffusion. Earlier studies claiming velocities of $\sim$10~km~s$^{-1}$ for young stars were probably affected by not so accurate measurements and their influence in the computation of velocities (e.g., \citealt{Mihalas1981} or \citealt{Tetzlaff2011}).

\subsection{Spatial distribution of the runaway stars} \label{Sec:SpatialDistribution}

The galactocentric $XY$ coordinates of the runaway stars of the GOSC- and BeSS-Gaia DR3 catalogs were shown in Figs.~\ref{Fig:GOSC_XY}~and~\ref{Fig:BeSS_XY}, respectively. As noted in Sec.~\ref{Sec:Runaways_Found},
there is no preferred spatial distribution for the runaways as there is for field stars, but there are no O-type runaway stars close to the Sun, while Be runaways are clustered around the solar neighborhood. To explain this behavior, two factors are named. On the one hand, because the brightest stars with $G<6$ were filtered, nearly all O-type stars were removed around the solar neighborhood, while this was not the case for the Be stars. On the other hand, the closer stars have smaller uncertainties and are therefore easier to be classified as runaways if they have a significant velocity. These are the reasons for the many runaway Be stars close to the Sun, while this is not the case for O stars.

The distribution of runaway stars might not trace the spiral arms because massive runaway stars are moving with high velocities and have been expelled from their birthplace \citep{Blaauw1993,MA2018}. Therefore, these stars can be found outside the spiral arms and are in particular more spread out along the $Z$-axis or the Galactic latitude $b$ than the field stars. This is shown for the $Z$-axis in Figs.~\ref{Fig:GOSC_Z}~and~\ref{Fig:BeSS_Z}. Field stars are mostly located close to the Galactic plane, while runaway stars are more distributed along the $Z$-axis. A similar effect is shown for the Galactic latitude $b$ in Figs.~\ref{Fig:GOSC_Aitoff}~and~\ref{Fig:BeSS_Aitoff}. The dispersions we obtained for $Z$ and $b$ for runaways (see Table~\ref{Tab:FitsAfterClippingRun}) are significantly larger
than for field stars (see Table~\ref{Tab:FitsAfterClippingField}). 
For the GOSC-\textit{Gaia}~DR3 catalog, the dispersion in $Z$ for the runaways is about three times larger than for the field stars. This factor is about six for the BeSS-\textit{Gaia}~DR3 catalog. For $b,$ these numbers are about two and about three, respectively. The dispersion differences in $Z$ and $b$ for the O and Be runaways are larger for Be runaways. This is explained by the fact that the Be stars are older, which allowed them to wander farther from the Galactic plane.

We note that the mean and standard deviation in $Z$ and $b$ were also obtained for stars up to 2~kpc only. For the field stars (Table~\ref{Tab:FitsAfterClippingField}), there are no significant differences with the complete catalog results. This indicates that our results are stable and that for the BeSS-\textit{Gaia}~DR3 catalog, they are not biased by the overdensity toward $XY = (13,5)$~kpc that extends from $Z\simeq0$~kpc close to the Sun to $Z\simeq0.4$~kpc away from it. In the case of the runaway stars (see Table~\ref{Tab:FitsAfterClippingRun}), the dispersions in $b$ do not change significantly, while there is a small decrease in the dispersions in $Z$. This last issue arises because runaway stars with higher values of $|Z|$ are rare and also located at larger $XY$ distances from the Sun, which effectively removed them with our $d<2$~kpc filter.

\subsection{Comparison with previous works} \label{Sec:Comparison}

For each of our runaways, we searched in the literature whether they had already been identified as runaways by other authors. As presented in Sect.~\ref{Sec:Runaways_Found}, we found 42 and 47 runaway stars with no previous identifications for the GOSC- and the BeSS-\textit{Gaia}~DR3 catalogs, respectively. We recovered 64 and 22 known runaway stars for the two catalogs, respectively. The corresponding publication references are listed in the column Run. Ref. of Tables~\ref{Tab:GOSC_Runaways}~and~\ref{Tab:BeSS_Runaways}. Here we mention the most common publications for our runaways. Considering both catalogs, we recovered 15, 41, 23, and 12 runaway stars that were previously identified by \cite{Tetzlaff2011}, \cite{MA2018}, \cite{Kobulnicky2022}, and \cite{Boubert2018}, respectively. We note that \cite{Kobulnicky2022} only published the names of 25 out of the 102 GOSC runaways they identified. Therefore, we cannot compare our results with their complete list. We show a summary of the samples, methods, thresholds and percentage of runaways published in these works in Table 6, together with the results obtained in this work.

\renewcommand{\arraystretch}{1.1}
\begin{table}[t!]
\caption{Stars and runaway stars as a function of spectral type.}
\label{Tab:SpectralType}      
\centering                          
\begin{tabular}{lcccc}
\hline \hline \vspace{-3mm}\\
Spectral Type  & 
Stars          &
\multicolumn{2}{c}{Runaway Stars} \vspace{-2mm} \\
\vspace{-2mm}
          &       & \multicolumn{2}{c}{-----------------------} 
\vspace{0mm}\\
          &  \#   & \#   & \%      \\
\hline \vspace{-3mm}\\
O2--O7    &  199  &  50  &   25.1  \\
O8--O9    &  194  &  46  &   23.7  \\
B0e--B3e  &  585  &  36  &  ~~6.2  \\
B4e--B9e  &  482  &  23  &  ~~4.8  \\
\hline
O8--B1e   &  420  &  62  &  14.8   \\
\hline
\end{tabular}
\end{table}

\renewcommand{\arraystretch}{1.1}
\begin{table*}
\caption{Main characteristics of works on all-sky searches of runaways among young stars discussed in Sec.\ref{Sec:Comparison}.}
\label{Tab:Publications}      
\centering                          
\begin{tabular}{l@{~~~~}c@{~~~~}c@{~~~~}c@{~~~~}c}
\hline \hline \vspace{-3mm}\\
Sample  & 
Method  &
Thresholds  &
Runaways  &
References  \\
\hline \vspace{-3mm}\\
\textit{Hipparcos} and $v_{\rm r}$; age $\leq$50~Myr; $d<3$~kpc  &  3D velocities  &  28~km~s$^{-1}$  & $\sim$27\% &  \citet{Tetzlaff2011} \\
GOSC and \textit{Gaia} DR1  &  2D proper motions     &  None  & $\sim$5.7\% &  \citet{MA2018} \\
GOSC and \textit{Gaia} EDR3 &  2D velocities         &  25~km~s$^{-1}$  & $\sim$22\% &  \citet{Kobulnicky2022} \\
GOSC and \textit{Gaia} DR3  &  2D velocities         &  None  & 25.4\% &  This work \\
GOSC and \textit{Gaia} DR3  &  3D velocities (sim.)  &  None  & 30.0\% &  This work \\
\hline
BeSS, \textit{Gaia} DR1 and LAMOST; B\&G (2001)  &  3D velocities  &  None  & $\sim$13\% &  \citet{Boubert2018} \\
BeSS and \textit{Gaia} DR3  &  2D velocities  &  None  & ~5.2\% &  This work \\
BeSS and \textit{Gaia} DR3  &  3D velocities (sim.)  &  None  & ~6.7\% &  This work \\
\hline
\end{tabular}
\tablefoot{ B\&G (2001) stands for \cite{Berger2001}. 3D velocities (sim.) stands for simulated 3D velocities (see Appendix \ref{Sec:App_Simulations}).}
\end{table*}

\cite{Tetzlaff2011} searched for young ($\leq$50~Myr) runaways around 3~kpc from the Sun in three dimensions using \textit{Hipparcos} data and radial velocities from the literature. They found a runaway percentage of $\sim$27\%. The limit in age used by these authors includes basically our first three bins of Table~\ref{Tab:SpectralType}, for which we would obtain $\sim$13.5\% of runaways, but using a 2D method. Although the samples are not the same and might be biased, we seem to recover only roughly 50\% of the runaway stars because we did not use radial velocities. Therefore, the percentages shown in Table~\ref{Tab:SpectralType} should probably be multiplied by a factor $\sim$2 to obtain the real percentages of runaways, which would imply percentages above 50\% for O-type stars. Because this percentage is quite high, we performed simulations of runaway stars using a 3D velocity method that reproduced our 2D results. We obtained a 3D percentage of $\sim$30\% for O-type stars compared to the 2D percentage of 25.4\% (see Appendix~\ref{Sec:App_Simulations}).

Our GOSC-\textit{Gaia}~DR3 runaway percentage of 25.4\% (see Sect.~\ref{Sec:Runaways_Found}) is significantly higher than that of \cite{MA2018}, who obtained 5.7\% for O stars. Differences lie slightly in the samples, but mainly in the \textit{Gaia} data, DR1 by them versus DR3 here, and especially in the method used, 2D proper motions by them versus 2D space velocities here. This prevented them from identifying stars with high peculiar velocity at large distances.

\cite{Kobulnicky2022} searched for runaway stars within GOSC using \textit{Gaia} EDR3 data and a 2D method, following the procedure of computing the velocities from \cite{Randall2015}. There are some notable differences between their approach and ours. \cite{Kobulnicky2022} did not use the zeropoint correction to the parallax, and most notably, the larger external parallax uncertainties proposed by \cite{MA2022}. Therefore, their results could be too optimistic in this sense. We also used a different rotation curve and solar motion values to compute the velocities. Although they used a Monte Carlo analysis to estimate the final uncertainties, they considered as input uncertainties those in parallax, proper motions, and solar motion, while we also considered the uncertainty in the galactocentric distance of the Sun and in the Galactic rotation curve (plus the small coordinate uncertainties). Finally, they classified as runaway stars stars with a peculiar 2D velocity above a threshold 25~km~s$^{-1}$. Of the 25 GOSC runaways with published names, we recover only 23, mainly because of our larger but more realistic external parallax uncertainties. We also note that for these stars, we obtain slightly lower velocities than theirs on average because of the parallax correction we applied (the use of different Galactic rotation curves and solar motion velocities had practically no influence). The total number and percentage of runaways obtained by these authors are very similar to ours: 102 by them versus 106 here, that is, $\sim$22\% versus $\sim$25\%. However, the final samples are different: They considered as runaway stars those with higher velocities because of their 25~km~s$^{-1}$ threshold, although their uncertainties are underestimated, while we used more realistic uncertainties, but had a threshold of 16~km~s$^{-1}$ that was derived from the dispersion of the field stars and the individual uncertainties. This means that our sample contains fewer false positives related to uncertainties, but includes stars that were classified as walkaways by other authors.

The above works are mainly concerned with runaway O stars. Although studies of B runaway stars can be found in the literature \citep{Hoogerwerf2001,Tetzlaff2011}, those dealing specifically with Be stars are less common. \cite{Boubert2018} exclusively searched for Be runaway stars within the BeSS and LAMOST \citep{Hou2016} catalogs and the stars of \cite{Berger2001} that fulfill several quality cuts, using \textit{Gaia} DR1 data and radial velocities from the literature (3D method). To classify the stars as runaways, they used a Bayesian method in which they compared the hypothesis that every star in their catalog is a runaway with the null hypothesis that it is a member of the Milky Way disk using priors. Within their sample of 632 Be stars, they found 13.1\% runaway stars, representing 83 objects. We recovered 12 of them in our BeSS-\textit{Gaia}~DR3 runaways. Their percentage of runaways of 13.1\% is considerably higher than the one obtained here of 5.2\% and those found in the literature for B stars (see Sect.~\ref{sec:SpectralTypeRunawaysDiscussion}). This discrepancy of a factor $\sim$2.5 could be understood considering that they used a 3D Bayesian method and \textit{Gaia} DR1 astrometric data. Again, we performed simulations of runaway stars using a 3D velocity method that reproduced our 2D results and obtained a 3D percentage of $\sim$6.7\% for Be stars compared to the 2D percentage of 5.2\% (see Appendix~\ref{Sec:App_Simulations}). \cite{Berger2001} also analyzed Be runaway stars with a 3D method and found $\sim$6.7\% runaway stars.

\subsection{Runaways as a function of the spectral type} \label{sec:SpectralTypeRunawaysDiscussion}

The percentage of runaways as a function of the spectral type that was presented in Sect.~\ref{sec:SpectralTypeRunawaysResults} decreases as follows: 25.1\%, 23.7\%, 6.2\%, and 4.8\% for O2$-$O7, O8$-$O9, B0e$-$B3e, and B4e$-$B9e, respectively. This trend was already noted by \cite{Blaauw1961}, who showed from 3D velocities that the runaway frequency as a function of the spectral type drops sharply from roughly 20\% among O-type stars to 2.5\% among B0--B0.5 stars, and it is even lower, 1.5\%, among B1--B5 stars. \cite{GiesAndBolton1986} obtained lower limits and estimated similar percentages of OB runaway stars, while \cite{Stone1979} increased the percentage of runaway O stars to $\sim$49\%. More recently, \cite{Tetzlaff2011} found that $\sim$27\% of the young stars are runaways, and \cite{Boubert2018} provided 13.1\% runaways in a sample of Be stars. The scatter among the results is therefore considerable. It mainly falls on the different input data, the method (2D or 3D), and the existence or absence of a velocity threshold to classify the stars as runaways. However, there is agreement in the sense that the percentage of runaway O stars is significantly higher than for B or Be stars.

In this context, two possible scenarios might explain the origin of runaway stars: the dynamical ejection scenario (DES; \citealt{Poveda1967}), and the binary supernova scenario (BSS; \citealt{Blaauw1961}). \cite{DorigoJones2020} compared these two scenarios for OB runaway stars in the SMC. They reported that the DES scenario often results in faster, more massive runaways, while the BSS scenario results in lower-velocity ejections of lower-mass stars. Their results for the SMC show that the dynamical mechanism, which favors massive runaways, dominates by a factor of 2--3. The GOSC-\textit{Gaia}~DR3 stars have higher velocities in general than those in BeSS-\textit{Gaia}~DR3 (see Figs.~\ref{Fig:GOSC_2D}~and~\ref{Fig:BeSS_2D}), which agrees with the scenarios discussed above. Moreover, we obtain a factor of $\sim$5 between the percentage of runaway stars among O-type stars versus Be-type stars in the Galaxy. This reinforces the dominance of the DES scenario versus the BSS one discussed by \cite{DorigoJones2020}.

\subsection{High-mass X-ray and gamma-ray binaries}

We cross-checked our catalogs of runaway stars with the catalog of High Mass X-ray Binaries (HMXBs) by \cite{Fortin2023}, which also contains gamma-ray binaries. For the GOSC-\textit{Gaia}~DR3 runaways, we found the following five objects in common with our names (their names) ordered by decreasing values of the $E$ parameter: \object{V479~Sct} (\object{LS~5039}), \object{GP~Vel} (\object{Vela~X-1}), \object{HD~153919} (\object{4U~1700$-$377}), \object{LM~Vel} (\object{IGR~J08408$-$4503}), \object{and Cygnus~X-1}. For the BeSS-\textit{Gaia}~DR3 runaways, we found the following two objects in common: \object{GSC~03588-00834} (\object{SAX~J2103.5+4545}) and \object{BQ~Cam} (\object{V~0332+53}). \object{LM~Vel} and \object{GSC~03588-00834} are identified here as runaway HMXBs for the first time.

Of all these HMXBs, we would like to comment on the particular case of \object{Cygnus~X-1}. We obtained a value of $E=1.10$, which is close to our threshold criterion to classify stars as runaways. The study by \cite{Mirabel2003} revealed that this object is moving with a space velocity of only $9\pm2$~km~s$^{-1}$ with respect to \object{Cyg~OB3}. They used this information to justify that \object{Cygnus~X-1} is a member of \object{Cyg~OB3} association and has a low space velocity with respect to its environment. However, the stars of \object{Cyg~OB3} are already moving with a space velocity of $\sim$22 km~s$^{-1}$ with respect to the mean Galactic rotation curve and basically toward the Galactic center \citep{Rao2020}, which in this case nearly coincides with the $V_{\text{TAN}}$ direction. Therefore, the slight additional velocity of \object{Cygnus~X-1} is enough for us to classify it as a runaway object, although it would not be classified as such considering the space velocity of the association. This is a limitation of our method, which could affect some stars with values of $E\gtrsim1$.

On the other hand, we are also interested in searching for new gamma-ray binaries. Only ten such systems are currently known according to \cite{Bordas2023}. Four of the eight systems with good spectral type classification for the massive stars are O stars and four are Be stars. Only the O star \object{LS~5039} is listed in the current version of the GOSC catalog, and we classify it as a runaway, as was already known \citep{Ribo2002,Moldon2012}. Three Be gamma-ray binaries are listed in the current version of the BeSS catalog: \object{PSR~B1259$-$63}, \object{LS~I~+61~303}, \object{and HESS~J0632+057}. Of these, only \object{PSR~B1259$-$63} is a runaway, but mainly in the radial direction, in which we are blind. Therefore, none of them was classified as a runaway by our method, as expected. The detection of \object{LS~5039} reveals that a careful inspection of our runaway stars, particularly among O-type stars, using multiwavelength catalogs could reveal new gamma-ray binaries.

\section{Summary and conclusions} \label{Sec:Conclusions}

Using \textit{Gaia} DR3 data, we have searched for Galactic massive runaway stars of O and Be spectral type. We cross-matched the GOSC and the BeSS catalog with \textit{Gaia} DR3 and obtained 417 and 1335 stars, respectively. We presented a 2D method to classify the stars as runaways if they had a significant velocity at a 3-sigma confidence level with respect to the field stars, which follow the Galactic rotation curve. The main conclusions of this work are summarized below. 

\begin{itemize}

    \item For O-type field stars, we obtained a velocity dispersion perpendicular to the Galactic plane of $\sim$5~km~s$^{-1}$ and a parallel dispersion of $\sim$7~km~s$^{-1}$. This increase is due to parallax uncertainties and the use of the Galactic rotation curve for distant stars.    

    \item For Be-type field stars, we obtained a velocity dispersion perpendicular to the Galactic plane of $\sim$5~km~s$^{-1}$ and a parallel dispersion of $\sim$9~km~s$^{-1}$. This increase is dominated by the Galactic velocity diffusion allowed by their older ages.

    \item We found 106 O runaway stars, representing 25.4\% of our O-type star catalog. Forty-two of them were not previously identified as runaways.
    
    \item We found 69 Be runaway stars, representing 5.2\% of our Be-type star catalog. Forty-seven of them were not previously identified as runaways.

    \item Considering O- and Be-type stars, the dispersion of runaways is about three to six and about two to three times higher in $Z$ and $b$ than that of field stars, respectively. This is explained by the ejections they underwent when they became runaways.
     
    \item The percentage of runaways decreases drastically from O- to Be-type stars, but it decreases gradually within each type: $\sim$25\%, $\sim$24\%, $\sim$6\%, and $\sim$5\% for O2$-$O7, O8$-$O9, B0e$-$B3e, and B4e$-$B9e, respectively. 

    \item Comparison with previous works using 3D methods suggests that these percentages could be $\sim$2 and $\sim$2.5 times higher for O- and Be-type stars, respectively. However, 3D simulations for our catalogs reveal that our 2D percentages could increase only up to $\sim$30\% and $\sim$6.7\%, respectively, implying factors of $\sim$1.2 and $\sim$1.3, respectively.

    \item We find higher velocities for O-type runaways than for Be-type runaways. This agrees with previous works. We find a factor of $\sim$5 between the percentage of runaway stars among O-type stars versus Be-type stars. Both facts underline that the DES scenario is more likely than the BSS scenario.
    
    \item Seven of our runaway stars are HMXBs, two of which are identified as new runaways in this work, and one is a gamma-ray binary. This opens the door to identifying new high-energy systems among our runaways by conducting detailed multiwavelength searches.

\end{itemize}

The upcoming \textit{Gaia} data releases, containing significantly improved parallaxes, together with the use of radial velocities, could lead us to discover more runaway stars among our input samples.

\begin{acknowledgements}

      We thank the anonymous referee for useful suggestions and comments that helped to improve the content of the manuscript.
      We would like to thank the members of the \textit{Gaia} team at Universitat de Barcelona for many useful discussions.
      We thank J. Maíz Apellániz for valuable discussions.
      We acknowledge financial support from the State Agency for Research of the Spanish Ministry of Science and Innovation under grants PID2019-105510GB-C31/AEI/10.13039/501100011033, PID2019-104114RB-C33/AEI/10.13039/501100011033, PID2022-136828NB-C41/AEI/10.13039/501100011033/ERDF/EU, and PID2022-138172NB-C43/AEI/10.13039/501100011033/ERDF/EU,   
      and through the Unit of Excellence María de Maeztu 2020-2023 award to the Institute of Cosmos Sciences (CEX2019-000918-M). We acknowledge financial support from Departament de Recerca i Universitats of Generalitat de Catalunya through grant 2021SGR00679. MC-C acknowledges the grant PRE2020-094140 funded by MCIN/AEI/10.13039/501100011033 and FSE/ESF funds.
      This work has made use of data from the European Space Agency (ESA) mission {\it Gaia} (\url{https://www.cosmos.esa.int/gaia}), processed by the {\it Gaia} Data Processing and Analysis Consortium (DPAC, \url{https://www.cosmos.esa.int/web/gaia/dpac/consortium}). Funding for the DPAC has been provided by national institutions, in particular the institutions participating in the {\it Gaia} Multilateral Agreement. This work has made use of the BeSS database, operated at LESIA, Observatoire de Meudon, France: \url{http://basebe.obspm.fr}.
      This research has made use of NASA’s Astrophysics Data System.
      This research has made use of the SIMBAD database, operated at CDS, Strasbourg, France.
\end{acknowledgements}

\bibliographystyle{aa}
\bibliography{mybibliography}

\begin{appendix}


\section{Quality cuts applied to the cross matches} \label{Sec:App_CutsTable}

Table~\ref{Tab:Cuts} shows the quality cuts we applied to the 598 and 1808 stars obtained after the cross matches of GOSC and BeSS with \textit{Gaia} DR3, respectively. Column 1 lists the quality cuts we used. Columns 2--3 show the number of GOSC stars (\#) affected by each quality cut and the corresponding percentage (\%) with respect to the total of 598 stars. Columns 4--5 show the same for BeSS stars, where percentages are given with respect to the total of 1808 stars. The last row of the table includes the resulting numbers after all the cuts were applied. This last row does not show the sum of the star numbers and percentages because some stars are affected by more than one quality cut. There are 423 GOSC stars and 1336 BeSS stars that survive after these quality cuts.

\renewcommand{\arraystretch}{1.1}
\begin{table}[h!]
\caption{Quality cuts applied to the 598 and 1808 stars obtained after the cross matches of GOSC and BeSS with \textit{Gaia} DR3, respectively.}
\label{Tab:Cuts}      
\centering                          
\resizebox{0.5\textwidth}{!}{
\begin{tabular}{@{}l@{~~}rr@{~~~~~}rr@{}}
\hline \hline \vspace{-3mm}\\
Quality Cut  &
\multicolumn{2}{c}{GOSC Stars}  &
\multicolumn{2}{c}{BeSS Stars} \\ 
                                                   & \#    & \%     & \#    & \% \\
\hline \vspace{-3mm}\\
Stars in pairs with same Gaia \texttt{source\_id}  &    6  &   1.0  &    2  &   0.1 \\
Non 5- or 6-parameter solution                     &    9  &   1.5  &   29  &   1.6 \\
$G<$~6                                             &   46  &   7.7  &  159  &   8.8 \\
\texttt{visibility\_periods\_used} $<$ 10          &    5  &   0.8  &   16  &   0.9 \\
$\sigma_{\rm ext}/\varpi_{\rm c}>0.2$              &   72  &  12.0  &  145  &   8.0 \\
Negative parallax                                  &    5  &   0.8  &   18  &   1.0 \\
RUWE $>$ 1.35                                      &  116  &  19.4  &  282  &  15.6 \\
\hline
All cuts & 175 & 29.3 & 472 & 26.1\\
\hline
\end{tabular}
}
\end{table}

\newpage

\FloatBarrier


\section{$U_{\text{RSR}}, V_{\text{RSR}}, \text{and } W_{\text{RSR}}$ velocity distributions} \label{Sec:App_velocities}

\begin{figure}
    \centering
   \includegraphics[width=\hsize]{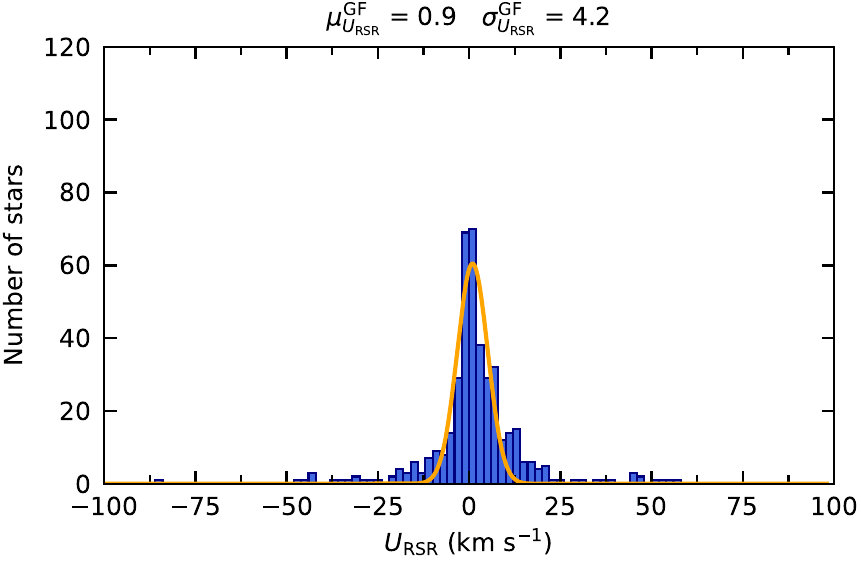}\vspace{3 mm}
    \includegraphics[width=\hsize]{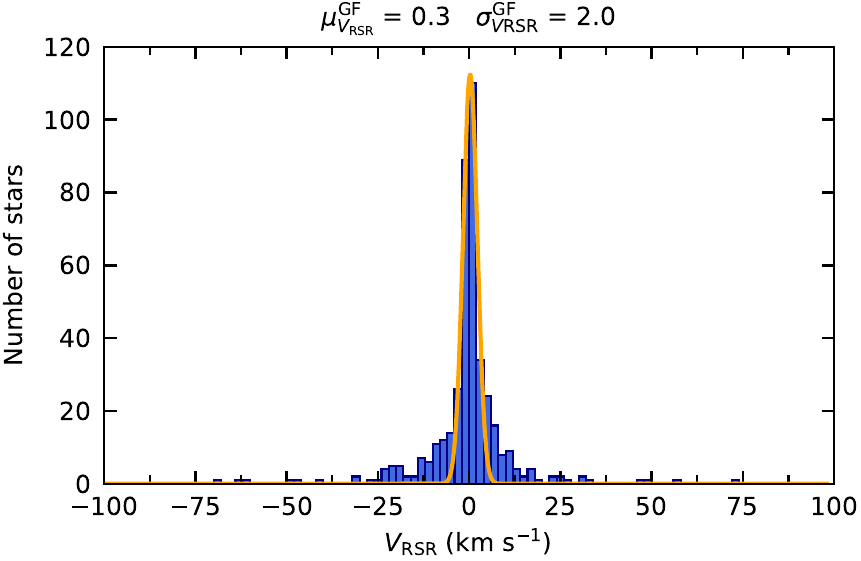}\vspace{3 mm}
  \includegraphics[width=\hsize]{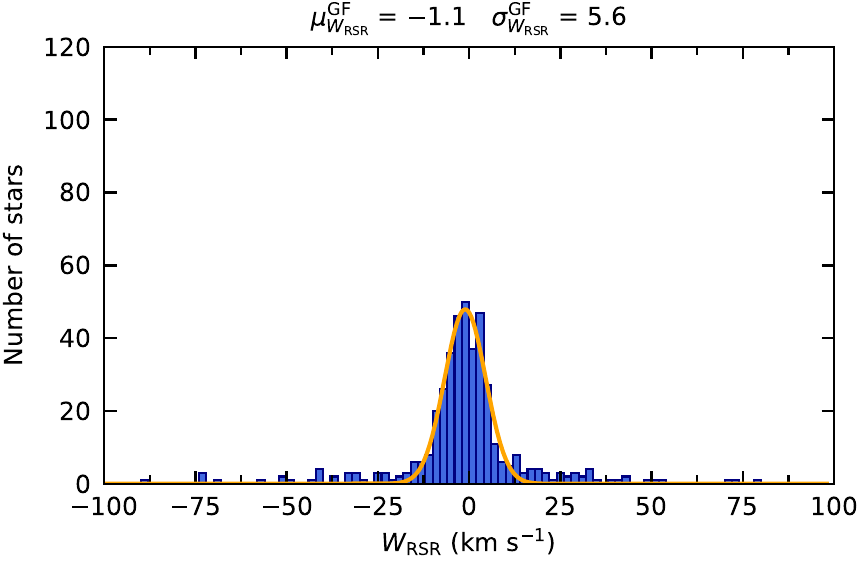}
    \caption{Distributions of the $U_{\text{RSR}}$, $V_{\text{RSR}}$ , and $W_{\text{RSR}}$ velocities for the GOSC-\textit{Gaia}~DR3 stars. The histograms have a bin size of 2~km~s$^{-1}$. The orange lines represent Gaussian functions fitted to the data, whose means and standard deviations in km~s$^{-1}$ are quoted above the panels. The abscissa ranges have been limited to $\pm 100$~km~s$^{-1}$.
    \textit{Top}: Histogram of the $U_{\text{RSR}}$ velocities. 
    \textit{Middle}: Histogram of the $V_{\text{RSR}}$ velocities.
    \textit{Bottom}: Histogram of the $W_{\text{RSR}}$ velocities.}
    \label{Fig:GOSC_RSRvels}
\end{figure}

\begin{figure}
    \centering 
    \includegraphics[width=\hsize]{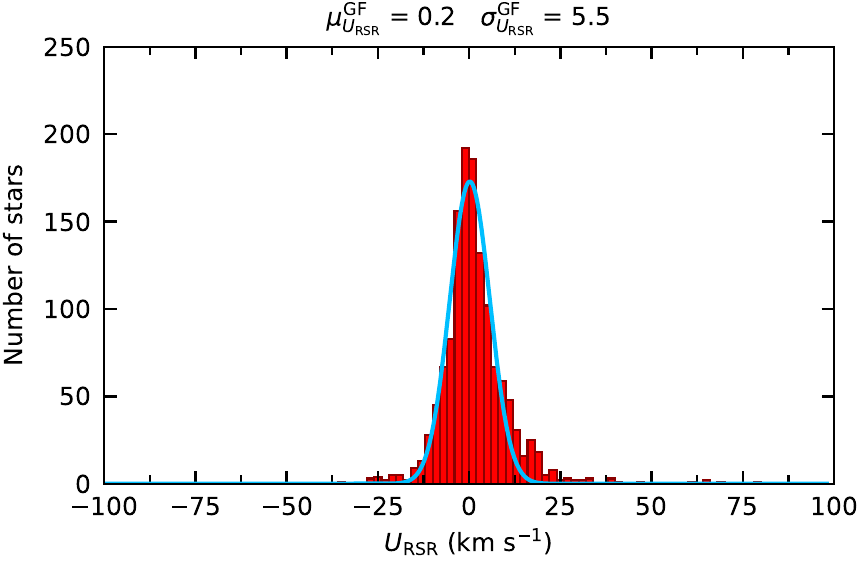}\vspace{3 mm}
    \includegraphics[width=\hsize]{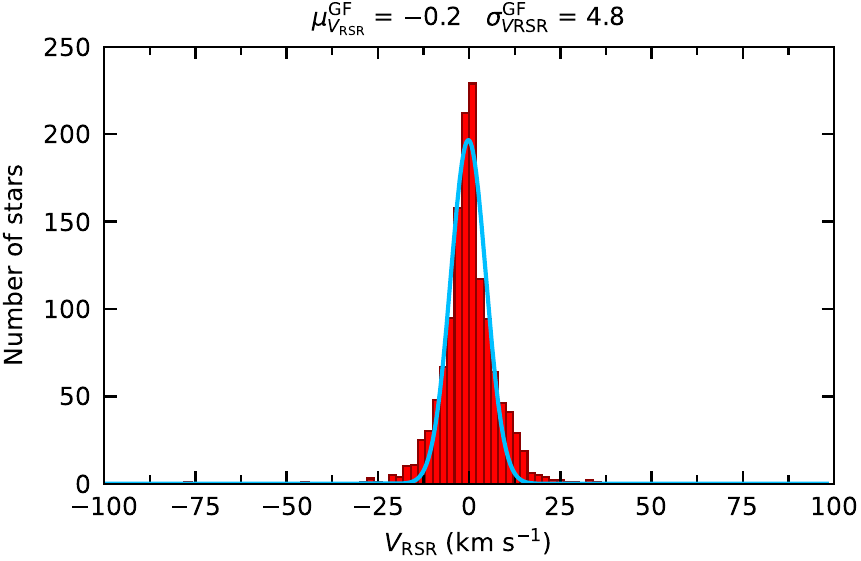}\vspace{3 mm}
    \includegraphics[width=\hsize]{Plots/BeSS/BeSS_WRSR.pdf}
    \caption{Same as Fig.~\ref{Fig:GOSC_RSRvels}, but for the BeSS-\textit{Gaia}~DR3 stars and with the Gaussian functions fitted to the data shown in blue.}
    \label{Fig:BeSS_RSRvels}
\end{figure}

The computed values of $U_{\text{RSR}}$, $V_{\text{RSR}}$, and $W_{\text{RSR}}$ are clustered around 0~km~s$^{-1}$. We produced histograms with a bin size of 2~km~s$^{-1}$ to better display the velocity distributions (as a compromise between resolution and significance). The histograms of the RSR velocities of the GOSC-\textit{Gaia}~DR3 stars are shown in Fig.~\ref{Fig:GOSC_RSRvels}. The top, middle, and bottom panels correspond to the $U_{\text{RSR}}$, $V_{\text{RSR}}$ , and $W_{\text{RSR}}$ velocities, respectively. The figures are limited to the $\pm 100$~km~s$^{-1}$ abscissa range for better visualization. For the $U_{\text{RSR}}$ and  $W_\text{RSR}$ velocities, two and three stars lie outside that range, respectively. The maximum absolute velocities of these three stars are $\lvert U_{\text{RSR}}\rvert = 177.1$~km~s$^{-1}$ and $\lvert W_{\text{RSR}}\rvert = 129.1$~km~s$^{-1}$. No stars lie outside the $V_\text{RSR}$ velocity range. The histograms show velocities clustered around zero, with some dispersion that approximately follows a Gaussian function. Some stars display much higher velocities. Gaussian fits are also shown in Fig.~\ref{Fig:GOSC_RSRvels}. The same histograms are shown for the BeSS-\textit{Gaia}~DR3 stars in Fig.~\ref{Fig:BeSS_RSRvels}. The figures are also limited to the $\pm 100$~km~s$^{-1}$ abscissa range for better visualization. Only one star lies outside that range in $U_{\text{RSR}}$ , with $\lvert U_{\text{RSR}}\rvert = 129.6$~km~s$^{-1}$, while no stars lie outside this range for the $V_{\text{RSR}}$ and  $W_\text{RSR}$ velocities.

For the GOSC- and BeSS-\textit{Gaia}~DR3 catalogs, the means and standard deviations of the Gaussian fits applied to the RSR velocity distributions using a bin size of 2~km~s$^{-1}$ are shown in Table~\ref{Tab:GuassianFits}. We note that the mean values are always closer to 0 than the bin size of 2~km~s$^{-1}$. The dispersions obtained in $U_{\text{RSR}}$, $V_{\text{RSR}}$ , and $W_{\text{RSR}}$ are about 5~km~s$^{-1}$ , except for $V_{\text{RSR}}$ for the GOSC-\textit{Gaia}~DR3 stars, which is significantly smaller. We emphasize that all velocities, but particularly $U_{\text{RSR}}$ and $V_{\text{RSR}}$, are affected by the lack of observational radial velocities. Our theoretical radial velocities decrease the dispersions of the distributions. For comparison, \cite{Tetzlaff2011} used proper motions from \textit{Hipparcos} and radial velocities from the literature for 7663 young stars to obtain dispersions in $U_{\text{RSR}}$, $V_{\text{RSR}}$ , and $W_{\text{RSR}}$ of 10.7, 10.7, and 5.3~km~s$^{-1}$, respectively. The dispersion in $W_{\text{RSR}}$ we derived is similar to theirs. Our dispersions in $U_{\text{RSR}}$ and $V_{\text{RSR}}$ are significantly smaller than theirs, as expected due to the use of theoretical radial velocities. 

\renewcommand{\arraystretch}{1.1}
\begin{table*}
\caption{Means and standard deviations of the Gaussian fits applied to the RSR velocity distributions using a bin size of 2~km~s$^{-1}$.}
\label{Tab:GuassianFits}
\centering
\begin{tabular}{lcr@{~~$\pm$~~}lr@{~~$\pm$~~}lr@{~~$\pm$~~}l}
\hline \hline \vspace{-3mm}\\
Catalog  &
Stars &
$\mu_{U_{\text{RSR}}}^{\text{GF}}$ & $\sigma_{U_{\text{RSR}}}^{\text{GF}}$  &
$\mu_{V_{\text{RSR}}}^{\text{GF}}$ & $\sigma_{V_{\text{RSR}}}^{\text{GF}}$  &
$\mu_{W_{\text{RSR}}}^{\text{GF}}$ & $\sigma_{W_{\text{RSR}}}^{\text{GF}}$  \\
 & \# &
\multicolumn{2}{c}{(km~s$^{-1}$)} &
\multicolumn{2}{c}{(km~s$^{-1}$)} &
\multicolumn{2}{c}{(km~s$^{-1}$)} 
\\
\hline \vspace{-3mm}\\
GOSC-\textit{Gaia}~DR3  &  ~~417  &  0.9 & 4.2  &     0.3 & 2.0  &  $-$1.0 & 5.6  \\
BeSS-\textit{Gaia}~DR3  &   1335  &  0.2 & 5.5  &  $-$0.2 & 4.8  &  $-$0.4 & 5.0  \\
\hline
\end{tabular}
\end{table*}

\FloatBarrier
~
\newpage

\section{Galactocentric $XZ$ and $YZ$ planes} \label{Sec:App_ZPlanes}

\begin{figure}[h]
    \centering
    \includegraphics[width=\hsize]{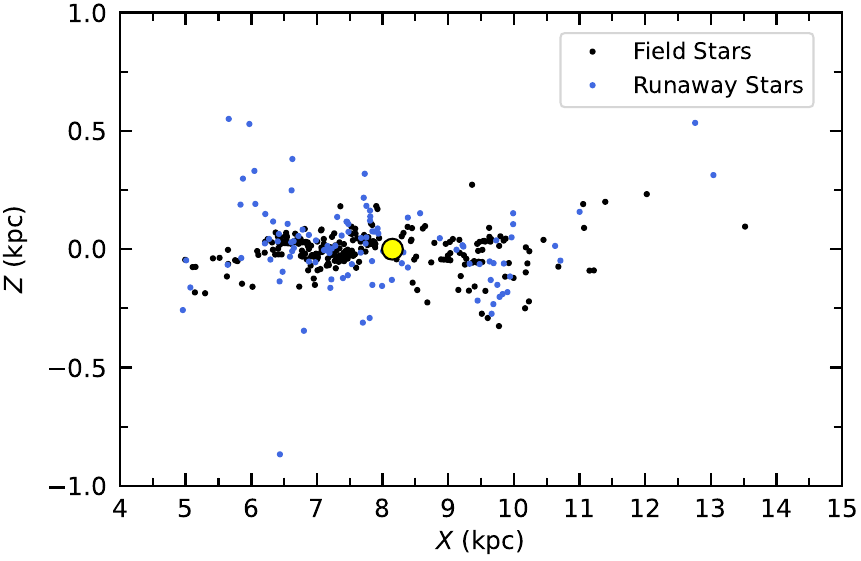}\vspace{3 mm}
    \includegraphics[width=\hsize]{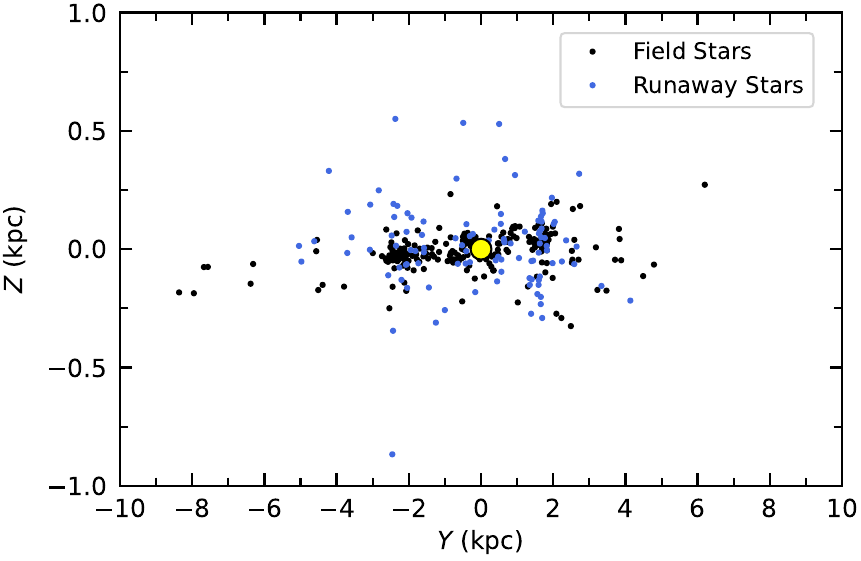}
    \caption{Galactocentric $XZ$ and $YZ$ coordinates for the stars of the GOSC-\textit{Gaia}~DR3 catalog. Field stars are depicted in black, and runaway stars are shown in blue. The position of the Sun is marked with a yellow circle. The ordinate axis is limited to $\pm1$~kpc and the axes have different scales. 
    \textit{Top}: Galactocentric $XZ$ coordinates.
    \textit{Bottom}: Galactocentric $YZ$ coordinates, but with a different scale in the abscissa axis.}
    \label{Fig:GOSC_Z}
\end{figure}

\begin{figure}[h]
    \centering
    \vspace{5.7cm}
    \includegraphics[width=\hsize]{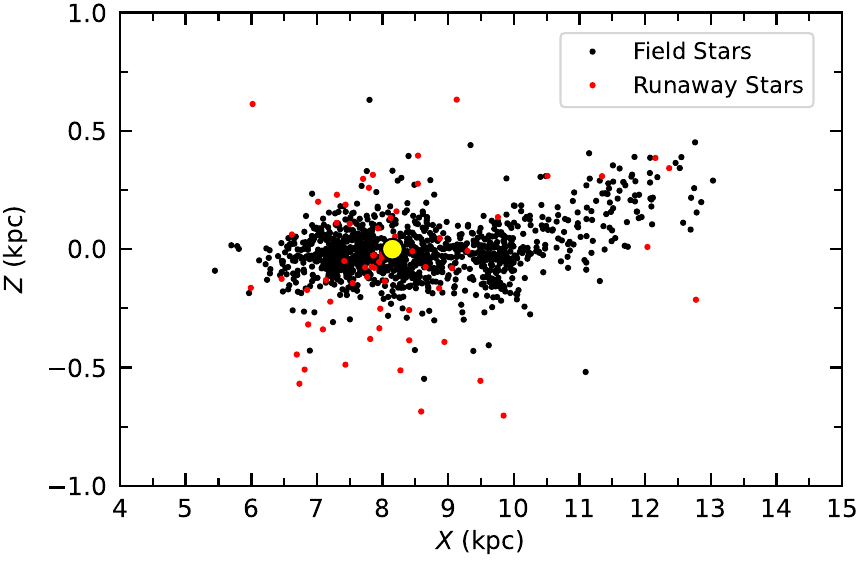}\vspace{3 mm}
    \includegraphics[width=\hsize]{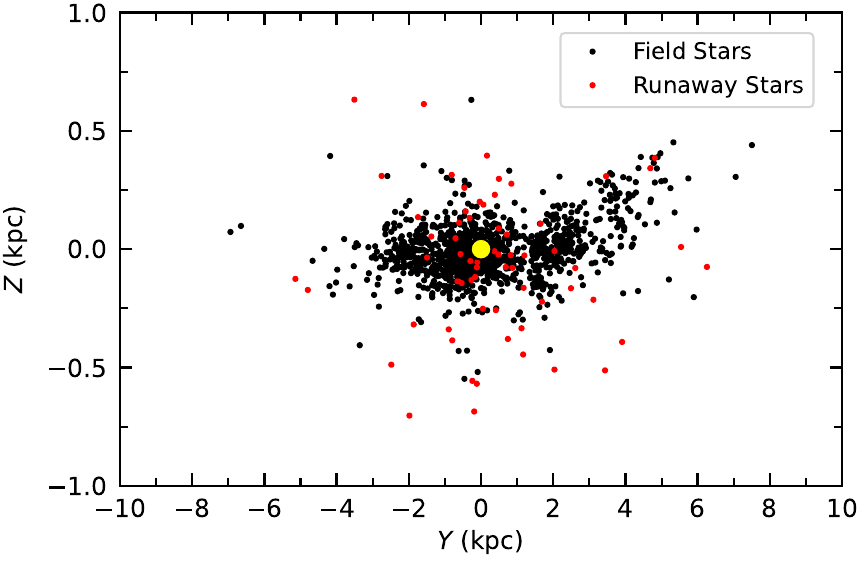}
    \caption{Same as Fig.~\ref{Fig:GOSC_Z} for the BeSS-\textit{Gaia}~DR3 stars. Field stars are depicted in black, and runaway stars are shown in red.}
    \label{Fig:BeSS_Z}
\end{figure}

Figure~\ref{Fig:GOSC_Z} shows the galactocentric $XZ$ (top) and $YZ$ (bottom) coordinates for the stars of the GOSC-\textit{Gaia}~DR3 catalog. Figure~\ref{Fig:BeSS_Z} shows the galactocentric $XZ$ (top) and $YZ$ (bottom) coordinates for the stars of the BeSS-\textit{Gaia}~DR3 catalog. In these figures, the $Z$ coordinate is limited to between $\pm 1$~kpc for a better visualization. Five stars in the BeSS-\textit{Gaia}~DR3 catalog lie outside that range, and their highest $|Z|$ value is $3.7$~kpc. The GOSC-\textit{Gaia}~DR3 stars are clustered around the Galactic plane with some dispersion (see main text). The BeSS-\textit{Gaia}~DR3 stars show a similar behavior, although the overdensity discussed in Sect.~\ref{Sec:Cross-match} toward $(X,Y)=(13,5)$ is located above the Galactic plane.

\FloatBarrier
~
\newpage

\section{Simulations of runaway stars using a 3D velocity method} \label{Sec:App_Simulations}

In this work, we used a 2D $\left(V_{\text{TAN}}, W_{\text{RSR}}\right)$ velocity method to compute the number of runaway stars in our catalogs. Here we report on the estimation of the number of runaway stars that we would detect using a 3D velocity method with simulated velocity distributions $\left(\widetilde{V}_{\text{LOS}}, \widetilde{V}_{\text{TAN}}, \widetilde{W}_{\text{RSR}}\right)$ that reproduce our 2D results for runaway stars.
We used the following approach.

\begin{enumerate}

    \item We used an exponential decay function  to simulate the modules of the 3D peculiar velocities for a population of \textit{N} stars. Specifically, we chose the python \texttt{random.exponential} function of the NumPy library, which draws samples from an exponential distribution following the equation $f(x,\beta)=(1/\beta)\exp{(-x/\beta)}$ with a given $\beta$ parameter. In addition, we multiplied the output values by a parameter $v_0$, in km~s$^{-1}$, which scales the exponential function to our desired range of velocities. We have three free parameters: the number of simulated stars \textit{N}, the $\beta$ parameter of the exponential function, and the velocity scale factor $v_0$.

    \item We assumed a 3D velocity distribution with spherical symmetry. To do this, we used a uniformly distributed random angle $\phi \in \left[0, 2\pi\right)$ and a uniformly distributed random function $\cos{\theta} \in \left[-1, 1\right]$, where $\theta \in \left[0, \pi\right]$.
    
    \item We computed the projections of the 3D simulated velocities $\left(\widetilde{V}_{\text{LOS}}, \widetilde{V}_{\text{TAN}}, \widetilde{W}_{\text{RSR}}\right)$ using the velocity modules and the $\phi$ and $\theta$ angles.
    
    \item For the velocity uncertainties of the stars, we assumed those corresponding to the 68\% percentile of the 2D uncertainty distributions of $\left(V_{\text{TAN}}, W_{\text{RSR}}\right)$. For the $\widetilde{V}_{\text{LOS}}$ component, we considered the same uncertainties as for $V_{\text{TAN}}$ because both are in the Galactic plane and are affected by similar uncertainties (corrected parallax, solar motion, and Galactic rotation curve). The velocity uncertainties we assumed are presented in the last three columns of Table~\ref{Tab:Simulations_Input}.

    \item We considered an underlying population of field stars. To do this, we took the standard deviations of $\left(V_{\text{TAN}}, W_{\text{RSR}}\right)$ obtained from the Gaussian fits, but centered at 0~km~s$^{-1}$ for simplicity (the differences are negligible). Again, for $\widetilde{V}_{\text{LOS}}$ , we considered the same standard deviation as for $V_{\text{TAN}}$. The means and standard deviations are presented in Table~\ref{Tab:Simulations_Input}.
  
    \item We classified 2D runaway stars as those with $E>1$ in Eq.~(\ref{Eq:Criteria}) using $\left(\widetilde{V}_{\text{TAN}}, \widetilde{W}_{\text{RSR}}\right)$ instead of $\left(V_{\text{TAN}}, W_{\text{RSR}}\right)$ and the input parameters presented in Table~\ref{Tab:Simulations_Input}. Similarly, we also classified 3D runaway stars adding the corresponding $\widetilde{V}_{\text{LOS}}$ term, $\left(\widetilde{V}_{\text{LOS}}, \widetilde{V}_{\text{TAN}}, \widetilde{W}_{\text{RSR}}\right)$, and input parameters to Eq.~(\ref{Eq:Criteria}).

    \item We conducted 10000 simulations to compute the average distributions of the simulated velocities $\left(\widetilde{V}_{\text{LOS}}, \widetilde{V}_{\text{TAN}}, \widetilde{W}_{\text{RSR}}\right)$, the mean numbers of 2D and 3D runaway stars, and the percentages with respect to the number of stars in our catalogs.

    \item We performed the following comparisons between the simulated population of 2D runaway stars found using $\left(\widetilde{V}_{\text{TAN}}, \widetilde{W}_{\text{RSR}}\right)$ and the population of 2D runaway stars found using $\left(V_{\text{TAN}}, W_{\text{RSR}}\right)$:
        
    \begin{itemize}

    \item The number of runaway stars in 2D.
    
    \item The standard deviations of the 2D velocities: $\sigma_{\widetilde{V}_{\text{TAN}}}$, $\sigma_{\widetilde{W}_{\text{RSR}}}$ (to be compared with  $\sigma_{V_{\text{TAN}}}$ and $\sigma_{W_{\text{RSR}}}$ of Table~\ref{Tab:FitsAfterClippingRun}).
        
    \item The histograms of $\widetilde{V}_{\text{TAN}}$, $\widetilde{W}_{\text{RSR}}$, and $\widetilde{V}_\text{PEC}^\text{2D}$ with respect to those of $V_{\text{TAN}}$, $W_{\text{RSR}}$, and $V_\text{PEC}^\text{2D}$.
        
    \item The 2D velocity plot $\widetilde{W}_{\text{RSR}}$ vs. $\widetilde{V}_{\text{TAN}}$ with respect to $W_{\text{RSR}}$ vs. $V_{\text{TAN}}$.
        
    \end{itemize}

    \item After these comparisons, we modified the three free parameters of the simulations in step~1 to reproduce the 2D runaway results obtained in this work as closely as possible, and we repeated the whole process until convergence was achieved. For the GOSC-\textit{Gaia}~DR3 catalog, convergence was achieved for the following parameters: $N=247$, $\beta=0.14$, $v_0=230$~km~s$^{-1}$. In the case of BeSS-\textit{Gaia}~DR3 catalog, the values were $N=328$, $\beta=0.11$, and $v_0=130$~km~s$^{-1}$.

\end{enumerate}

\renewcommand{\arraystretch}{1.1}
\begin{table*}
\caption{Input parameters for the simulations.}
\label{Tab:Simulations_Input}
\centering
\begin{tabular}{lccccccccc}
\hline \hline \vspace{-3mm}\\
Catalog  &  $\mu_{\widetilde{V}_{\text{LOS}}}^{\text{GF}}$  &  $\mu_{\widetilde{V}_{\text{TAN}}}^{\text{GF}}$  &  $\mu_{\widetilde{W}_{\text{RSR}}}^{\text{GF}}$ & 
$\sigma_{\widetilde{V}_\text{LOS}}^{\text{GF}}$  &  $\sigma_{\widetilde{V}_\text{TAN}}^{\text{GF}}$  &  $\sigma_{\widetilde{W}_\text{RSR}}^{\text{GF}}$  &
$\sigma_{\widetilde{V}_{\text{LOS},*}}$  &   
$\sigma_{\widetilde{V}_{\text{TAN},*}}$  &  
$\sigma_{\widetilde{W}_{\text{RSR},*}}$ \\
&
(km~s$^{-1}$) &
(km~s$^{-1}$) &
(km~s$^{-1}$) &
(km~s$^{-1}$) &
(km~s$^{-1}$) &
(km~s$^{-1}$) &
(km~s$^{-1}$) &
(km~s$^{-1}$) &
(km~s$^{-1}$) \\
\hline \vspace{-3mm}\\
GOSC-\textit{Gaia}~DR3  &  0.0  &  0.0  &  0.0  &  6.6  &  6.6  &  5.3  &  5.7  &  5.7  &  1.5 \\
BeSS-\textit{Gaia}~DR3  &  0.0  &  0.0  &  0.0  &  9.3  &  9.3  &  4.9  &  5.0  &  5.0  &  1.1 \\
\hline
\end{tabular}
\end{table*}

\renewcommand{\arraystretch}{1.1}
\begin{table*}
\caption{Output results of the simulations.}
\label{Tab:Simulations_Output}
\centering
\begin{tabular}{lcccc|ccccc}
\hline \hline \vspace{-3mm}\\
Catalog  &
\multicolumn{4}{c}{Results of Simulated Runaway Stars in 2D} &
\multicolumn{5}{|c}{Results of Simulated Runaway Stars in 3D} \\
 & \multicolumn{1}{c}{Number} & \multicolumn{1}{c}{Percentage} &
$\sigma_{\widetilde{V}_{\text{TAN}}}$ &
$\sigma_{\widetilde{W}_{\text{RSR}}}$ &
\multicolumn{1}{c}{Number} & \multicolumn{1}{c}{Percentage
} &
$\sigma_{\widetilde{V}_{\text{LOS}}}$ &
$\sigma_{\widetilde{V}_{\text{TAN}}}$ &
$\sigma_{\widetilde{W}_{\text{RSR}}}$ \\
 & \multicolumn{1}{c}{\#} & \multicolumn{1}{c}{\%} &
(km~s$^{-1}$) &
(km~s$^{-1}$) &
\multicolumn{1}{c}{\#} & \multicolumn{1}{c}{\%} &
(km~s$^{-1}$) &
(km~s$^{-1}$) &
(km~s$^{-1}$) \\
\hline \vspace{-3mm}\\
GOSC-\textit{Gaia}~DR3  &  106 &  25.4 &  39  &  39  &  125 &  30.0  &  36  &  36  &  36 \\
BeSS-\textit{Gaia}~DR3  & ~~69 & ~~5.2 &  34  &  23  &   90 & ~~6.7  &  31  &  31  &  21 \\
\hline
\end{tabular}
\end{table*}

\begin{figure}
    \centering
    \includegraphics[width=\hsize]{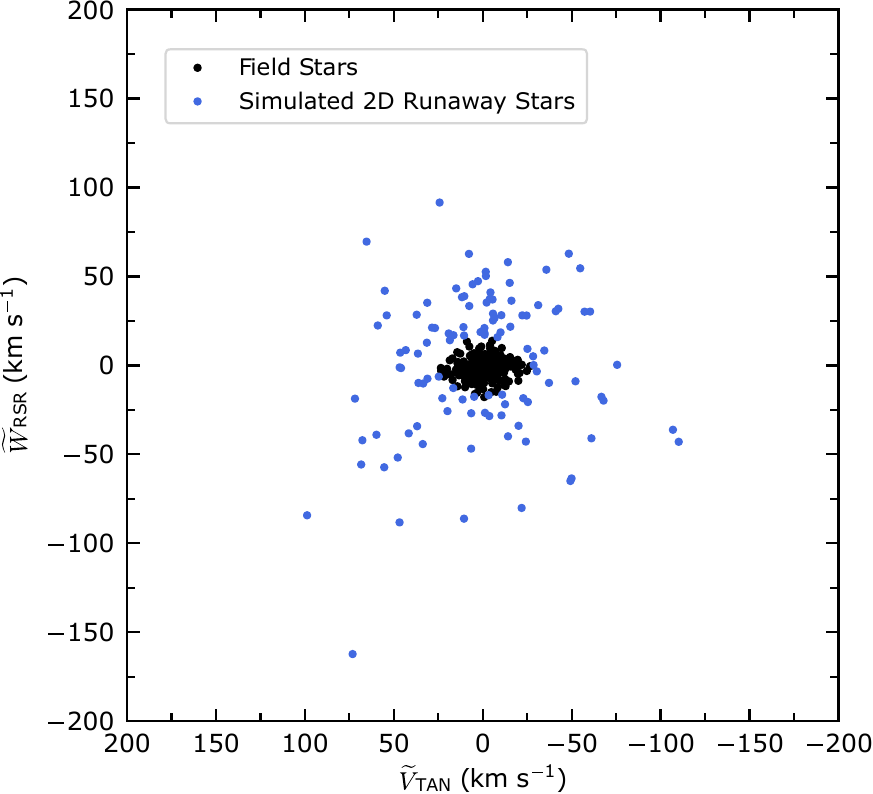}\vspace{3 mm}
    \includegraphics[width=\hsize]{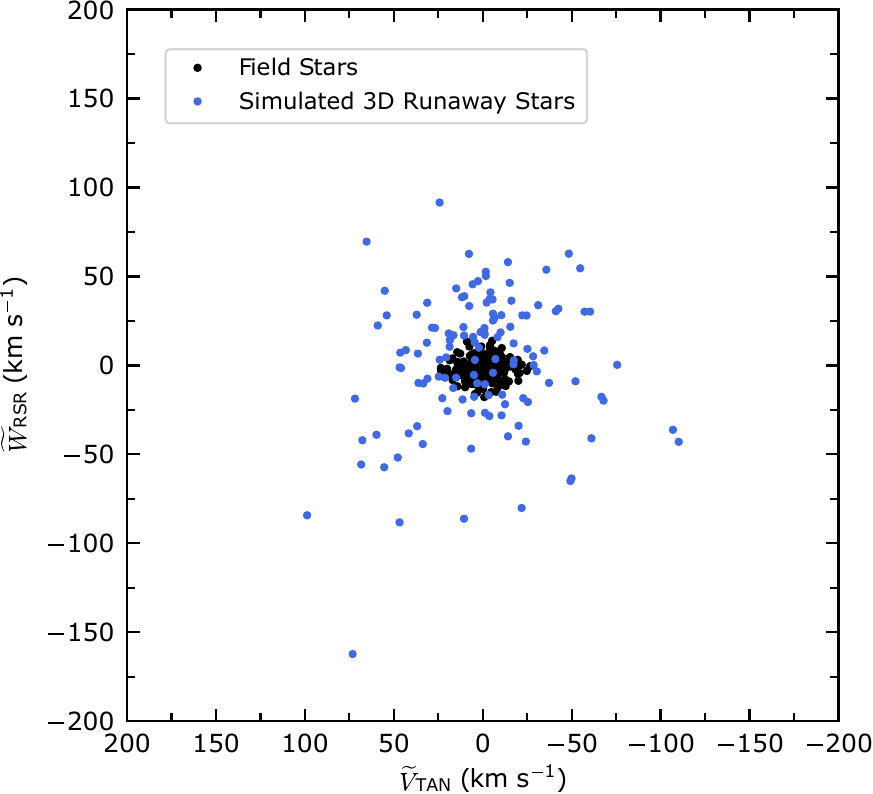}\vspace{3 mm}
    \caption{$\widetilde{W}_{\text{RSR}}$ as a function of $\widetilde{V}_{\text{TAN}}$ for the simulated runaway stars represented in blue for one of the 10000 simulations that match the GOSC-\textit{Gaia} DR3 results. The $\left(V_{\text{TAN}}, W_{\text{RSR}}\right)$ velocities of the field stars of the GOSC-\textit{Gaia} DR3 catalog are depicted in black.
    \textit{Top}: Simulated 2D runaway stars.
    \textit{Bottom}: Simulated 3D runaway stars with either positive or negative values of $\widetilde{V}_{\text{LOS}}$.}
    \label{Fig:GOSC_simulations}
\end{figure}

The results of our simulations for O-type stars are shown in Table~\ref{Tab:Simulations_Output}. In the 2D case, we accurately reproduce the number and percentage of runaway stars and the standard deviations of their velocities (see Table~\ref{Tab:FitsAfterClippingRun}). We show in the top panel of Fig.~\ref{Fig:GOSC_simulations} the 2D $\left(\widetilde{V}_{\text{TAN}}, \widetilde{W}_{\text{RSR}}\right)$ velocity distribution of the simulated runaway stars in blue for one of the 10000 simulations. The field stars of the GOSC-\textit{Gaia} DR3 catalog are plotted in black for reference. This figure resembles Fig.~\ref{Fig:GOSC_2D}. In the 3D case, we obtained 125 runaway stars, representing 30\% of the 417 stars in the GOSC-\textit{Gaia} DR3 catalog. The standard deviations of the runaway stars preserve the spherical symmetry, and are smaller than those in 2D because we can now detect stars as runaways with lower $\widetilde{V}_{\text{TAN}}$ and $\widetilde{W}_{\text{RSR}}$ velocities. This is clearly visible in the bottom panel of Fig.~\ref{Fig:GOSC_simulations}, which includes the simulated 3D runaway stars. In this figure, runaway stars lie in the region in which field stars were located in 2D because of the contribution of the $\widetilde{V}_{\text{LOS}}$  (for both positive and negative values of $\widetilde{V}_{\text{LOS}}$).

\begin{figure}
    \centering
    \includegraphics[width=\hsize]{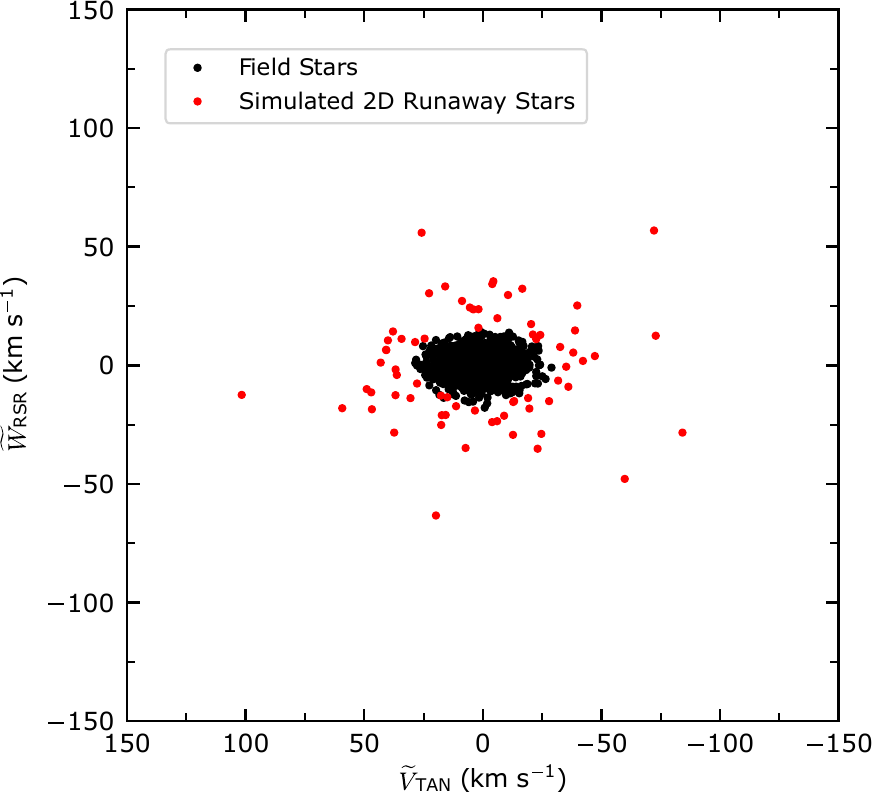}\vspace{3 mm}
    \includegraphics[width=\hsize]{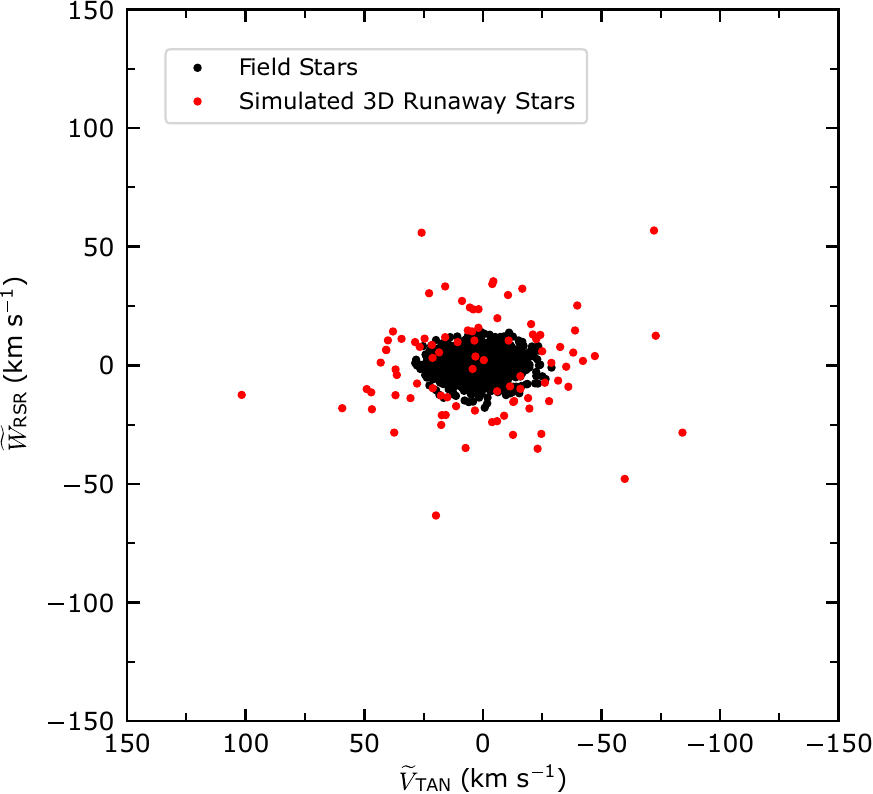}\vspace{3 mm}
    \caption{Same as Fig.~\ref{Fig:GOSC_simulations} for the BeSS-\textit{Gaia}~DR3 stars. Field stars are depicted in black, and runaway stars are shown in red.}
    \label{Fig:BeSS_simulations}
\end{figure}

For the Be-type stars, we applied the same process as described in steps 1 to 9 with the input parameters presented in Table~\ref{Tab:Simulations_Input}, but with a slight difference. In step~3, we multiplied the $\widetilde{V}_{\text{LOS}}$ and $\widetilde{V}_{\text{TAN}}$ distributions by a factor 1.5 to reproduce the BeSS-\textit{Gaia} DR3 velocity distributions because the standard deviation of $V_{\text{TAN}}$ obtained from the Gaussian fit is clearly higher than that of $W_{\text{RSR}}$ (see Table~\ref{Tab:FitsAfterClippingField}). We note that this can be understood as a result of Galactic velocity diffusion in the disk (see Sect.~\ref{Sec:Discussion_Velocities}). The results of our simulations for Be-type stars are shown in Table~\ref{Tab:Simulations_Output}. In the 2D case, the number, percentage, and the velocity standard deviations of the runaway stars were also reproduced (see Table~\ref{Tab:FitsAfterClippingRun}). We show in the top panel of Fig.~\ref{Fig:BeSS_simulations} the 2D $\left(\widetilde{V}_{\text{TAN}}, \widetilde{W}_{\text{RSR}}\right)$ velocity distribution of the simulated runaway stars in red for one of the 10000 simulations. The field stars of the BeSS-\textit{Gaia} DR3 catalog are shown in black for reference. This figure is similar to Fig.~\ref{Fig:BeSS_2D}. In the 3D case, we obtained 90 runaway stars, representing 6.7\% of the 1335 stars in the BeSS-\textit{Gaia} DR3 catalog. As expected due to the 1.5 factors used, the standard deviations of the runaway stars are equal for the $\widetilde{V}_{\text{LOS}}$ and $\widetilde{V}_{\text{TAN}}$ components, and smaller for the $\widetilde{W}_{\text{RSR}}$ component. They are also smaller than those in 2D. The 2D $\left(\widetilde{V}_{\text{TAN}}, \widetilde{W}_{\text{RSR}}\right)$ velocity distribution including the simulated runaway stars in 3D is shown in the bottom panel of Fig.~\ref{Fig:BeSS_simulations}, where we also have new runaways in the area of the 2D field stars because of the $\widetilde{V}_{\text{LOS}}$ contribution.

When the 2D and 3D results are compared, the number of runaways for O-type stars increased by $19\pm4$ from 2D to 3D (the uncertainty reflects the standard deviation of the increase in the 10000 simulations). This represents an increase from $\sim$25\% in 2D to $\sim$30\% in 3D, which corresponds to a factor of $\sim$1.2. For the Be stars, the increase in the number of runaways was $21\pm4$ from 2D to 3D. This represents an increase from $\sim$5.2\% in 2D to $\sim$6.7\% in 3D, corresponding to a factor of $\sim$1.3.

Finally, we note that we also used a power-law function to simulate the modules of the 3D peculiar velocities in step~1 of the simulations presented above and obtained similar results, although the exponential decay function reproduced the observational results better.

\end{appendix}

\end{document}